\documentclass{aart} 
\usepackage{subfiles} 

\usepackage{amsmath} 
\usepackage{amssymb}
\usepackage{amsthm}
\usepackage{amsxtra}
\usepackage{amscd}
\usepackage{mathabx}
\usepackage{bbm}
\usepackage{mathrsfs}
\usepackage{bm}
\usepackage{stmaryrd}
\usepackage{xspace} 
\usepackage{amsfonts}
\usepackage{latexsym}
\usepackage{graphicx} 
\usepackage{float}
\usepackage{caption} 
\usepackage{subcaption} 
\usepackage{comment} 
\usepackage{subfiles} 

\def\bfseries{\fontseries \bfdefault \selectfont \boldmath}
\def\bfdefault{b}

\usepackage{color}

\numberwithin{equation}{section}

\newcommand{\eq}[1]{\begin{equation}#1\end{equation}}
\newcommand{\eqs}[1]{\begin{equation*}#1\end{equation*}}
\newcommand{\al}[1]{\begin{align}#1\end{align}}

\renewcommand{\cal}{\mathcal} 

\newcommand{\dd}{\mathrm{d}} 
\newcommand{\deq}{\mathrel{\mathop:}=} 
\newcommand{\I}{\mathbbmss{1}} 
\renewcommand{\leq}{\leqslant} 
\renewcommand{\geq}{\geqslant} 

\renewcommand{\P}{\mathbb{P}}
\newcommand{\E}{\mathbb{E}}

\newcommand{\R}{\mathbb{R}}

\newcommand{\N}{\mathbb{N}}

\newcommand{\e}{{\varepsilon}}

\newcommand{\p}[1]{({#1})}

\newcommand{\pa}[1]{\left({#1}\right)}

\newcommand{\qa}[1]{\left[{#1}\right]}

\newcommand{\ha}[1]{\left\{{#1}\right\}}

\newcommand{\absa}[1]{\left\lvert #1 \right\rvert}

\newcommand{\floor}[1]{\lfloor #1 \rfloor}

\newcommand{\db}[1]{\llbracket{#1}\rrbracket}

\newcommand{\rhoM}{\rho_{\text{M}_n}}

\theoremstyle{plain} 
\newtheorem{theorem}{Theorem}[section]
\newtheorem*{theorem*}{Theorem}
\newtheorem{lemma}[theorem]{Lemma}
\newtheorem*{lemma*}{Lemma}

\newtheorem*{corollary*}{Corollary}

\newtheorem*{proposition*}{Proposition}
\newtheorem{definition}[theorem]{Definition}
\newtheorem*{definition*}{Definition}

\newtheorem*{example*}{Example}
\theoremstyle{remark} 
\newtheorem{remark}[theorem]{Remark}
\newtheorem*{remark*}{Remark}

\newtheorem*{remarks*}{Remarks}

\newtheorem*{notation*}{Notation}
\renewcommand{\qedsymbol}{$\blacksquare$} 

\title{Universality of Fixation Probabilities in\\Randomly Structured Populations}

\date{}

\author{Ben Adlam\textsuperscript{1,2} and Martin A. Nowak\textsuperscript{1,3,4}\\
Program for Evolutionary \mbox{Dynamics,\textsuperscript{1}} 
School of Engineering and Applied \mbox{Science,\textsuperscript{2}} \\
Department of \mbox{Mathematics,\textsuperscript{3}} and Department of Organismic and Evolutionary \mbox{Biology,\textsuperscript{4}} \\
Harvard University, Cambridge MA 02138, USA
} 

\def\biblio{\bibliography}

\begin{document}

\def\biblio{} 


\maketitle

\vspace{0.7cm}

\medskip

\begin{abstract}
{\sc Abstract:} The stage of evolution is the population of reproducing individuals. 
	The structure of the population is know to affect the dynamics and outcome of evolutionary processes, but analytical results for generic random structures have been lacking. 
	The most general result so far, the isothermal theorem, assumes the propensity for change in each position is exactly the same, but realistic biological structures are always subject to variation and noise. 
	We consider a population of finite size $n$ under constant selection whose structure is given by a wide variety of weighted, directed, random graphs; vertices represent individuals and edges interactions between individuals. 
	By establishing a robustness result for the isothermal theorem and using large deviation estimates to understand the typical structure of random graphs, we prove that for a generalization of the Erd\H{o}s-R\'{e}nyi model the fixation probability of an invading mutant is approximately the same as that of a mutant of equal fitness in a well-mixed population with high probability.
	Simulations of perturbed lattices, small-world networks, and scale-free networks behave similarly.
	We conjecture that the fixation probability in a well-mixed population, $(1-r^{-1})/(1-r^{-n})$,  is universal: for many random graph models, the fixation probability approaches the above function uniformly as the graphs become large.\\
\end{abstract}


\section*{}

In physics, a system exhibits universality when its macroscopic behavior is independent of the details of its microscopic interactions \cite{Deift2006}. Many physical models are conjectured as universal and long programs have been carried out to establish this mathematically \cite{Erdos2012, Borodin2012}. However such universality conjectures have been lacking in biological models.

It is well known that population structure can affect the behavior of evolutionary processes under both constant selection \cite{levin1974disturbance, levin1976population, durrett1994stochastic, Lieberman2005, Broom2008, Frean2013}, on which we focus here, and frequency dependent selection \cite{Sinervo1996, Kerr2002, Ohtsuki2006, Chen2013, Broom2013}. However, so far, deterministic and highly organized population structures have received the most attention \cite{Johst1999, Broom2009, Ispolatov2009, Diaz2013, Jamieson-Lane2013} while some populations are accurately modeled in this way \cite{Durrett1994, Hassell1994, Rainey2003, LeGac2010a, Allen2013}, often a random structure is far more appropriate to describe the irregularity of the real world \cite{Erdos1960, Watts1998, Barrat2000a, Nakamaru2004}. Random population structures have been considered numerically, but analytical results have been lacking \cite{Lieberman2005, Ohtsuki2006, Barbosa2010}.

The \emph{Moran process} considers a population of $n$ individuals, each of which is either wild-type or mutant with constant fitness 1 or $r$ respectively, undergoing reproduction and death \cite{Moran1962}. At each discrete time step an individual is chosen randomly for reproduction proportional to its fitness; another individual is chosen uniformly at random for death and is replaced by a new individual of the same phenotype as the reproducing individual. In the long run, the process has only two possible outcomes: the mutants fix and the wild-type dies out or the reverse. When a single mutant is introduced randomly into a homogenous, wild-type population, we call the probability of the first eventuality \emph{the fixation probability}. 

Fixation probabilities are of fundamental interest in evolutionary dynamics \cite{Patwa2008}. For a well-mixed population as described above, the fixation probability, denoted 
\eqs{\tag{1}
	\rhoM(r)=\frac{1-r^{-1}}{1-r^{-n}},
}
depends on $r$ and $n$ \cite{Maruyama1974a, Nowak2006}. Fixation probabilities also depend on population structure \cite{Barton1993, Whitlock2003}, which is modeled by running the process on a graph (a collection of $n$ vertices with edges between them) where vertices represent individuals and edges competition between individuals. Population structure forces reproducing individuals to replace only individuals with whom they are in competition, as described by the graph, and thus death is no longer uniformly at random but among only the reproducing individual's neighbors. See the Appendix for details.

With this enrichment of the model, the effects of population structure can be understood. Simple one-rooted population structures are able to repress selection and reduce evolution to a standstill, while intricate, star-like structures can amplify the intensity of selection to all but guarantee the fixation of mutants with arbitrarily slight fitness advantages \cite{Lieberman2005}. The former has been proposed as a model for understanding the necessity of hierarchical lineages of cells to reduce the likelihood of cancer initiation \cite{Nowak2003}. Some population structures have fixation probabilities which are given exactly by $\rhoM(r)$ and a fundamental result, called the isothermal theorem (stated precisely in Theorem \ref{thm: isothermal}), gives conditions for this \cite{Lieberman2005}. As a special case of these conditions are all symmetric population structures or graphs with undirected edges. More generally, a graph is called \emph{isothermal} if the sums of the outgoing and ingoing edge weights are the same for all subsets of the graph's vertices. This is our first hint of universality but it was not the first time certain quantities were observed as independent of population structure. Maruyama introduced geographical population structure by separating reproduction, which occurs within sub-populations, and migration, which occurs between sub-populations, and found that the fixation probability was the same as that of a well-mixed population structure \cite{Maruyama1974b}. In the framework of evolutionary graph theory, Maruyama's model would correspond to a symmetric graph. In this sense his finding is a special case of the isothermal theorem.

However, the assumptions of the isothermal theorem sit on a knife edge---when any small perturbation is made to the graph, the assumptions no longer hold and the original isothermal theorem is silent. In particular, it cannot be applied to directed, random graphs. We address these shortcomings in Section \ref{sec: robust isothermal}, where we strengthen the forward direction of the isothermal theorem by proving a deterministic statement: we weaken the theorem's assumptions to be only approximately true for a graph $G$ and show that the conclusion is still approximately true, that is, the fixation probability of a general graph $\rho_{G_{n}}(r)$ is approximately equal to $\rhoM(r)$. We call this the robust isothermal theorem ({\sc rit}).

\begin{theorem*}[Robust isothermal theorem]
	Fix $0\leq \e<1$. Let $G_{n}=(V_{n},W_{n})$ be a connected graph. If for all nonempty $ S\subsetneq V_{n}$ we have
	\eq{\tag{2}\label{eq: main assumption}
		\absa{\frac{w_{\text{O}}(S)}{w_{\text{I}}(S)}-1}\leq \e,
	}
	where $w_{\text{O}}(S)$ and $w_{\text{I}}(S)$ are the sums of the outgoing and ingoing edges respectively, then
	\eq{\tag{3}
		\sup_{r>0}\absa{\rhoM(r) - \rho_{G_{n}}(r)} \leq  \e.
	}
\end{theorem*}

\begin{figure}[H]
	\centering
	\begin{subfigure}{.24\textwidth}
		\centering
		\includegraphics[width=.9\linewidth]{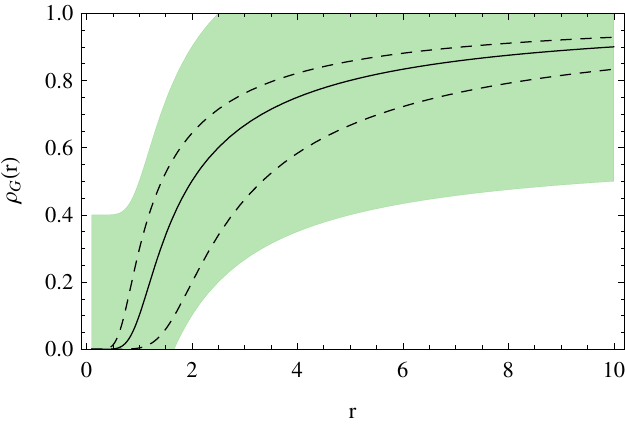}
	\end{subfigure}%
	\begin{subfigure}{.24\textwidth}
		\centering
		\includegraphics[width=.9\linewidth]{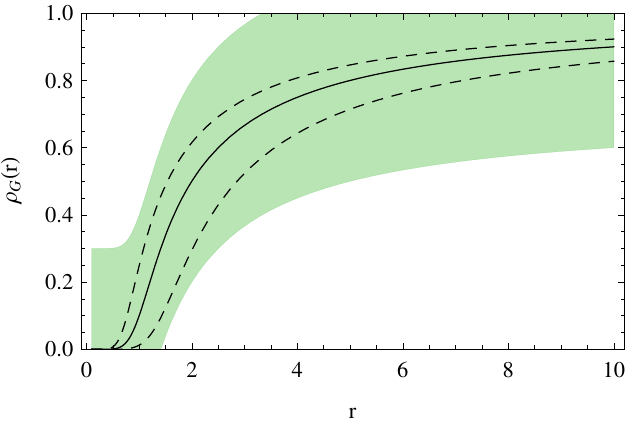}
	\end{subfigure}
	\begin{subfigure}{.24\textwidth}
		\centering
		\includegraphics[width=.9\linewidth]{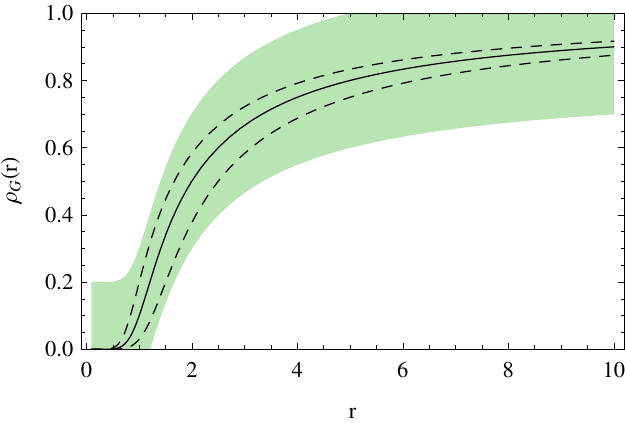}
	\end{subfigure}%
	\begin{subfigure}{.24\textwidth}
		\centering
		\includegraphics[width=.9\linewidth]{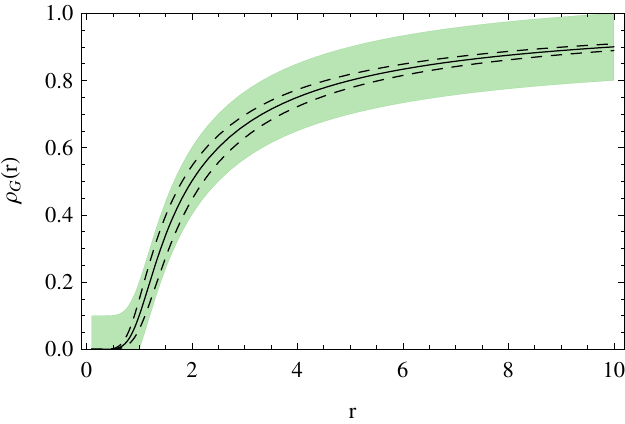}
	\end{subfigure}
	\\
	\begin{subfigure}{.03\textwidth}
		\raggedright
		\includegraphics[width=1\linewidth]{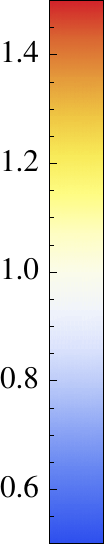}
	\end{subfigure}%
	\begin{subfigure}{.24\textwidth}
		\centering
		\includegraphics[width=.9\linewidth]{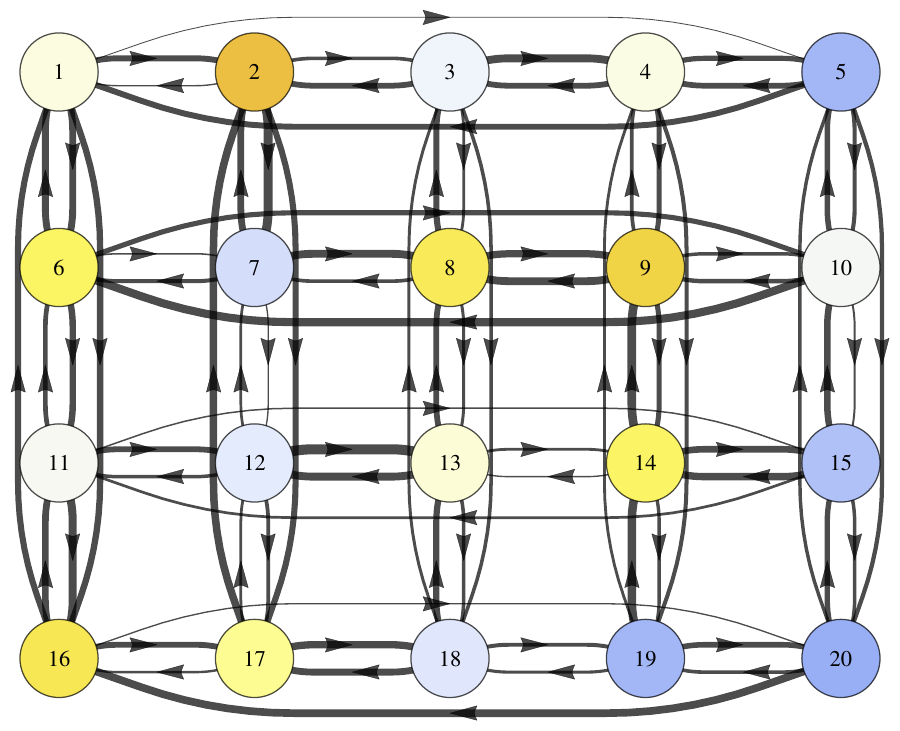}
	\end{subfigure}%
	\begin{subfigure}{.24\textwidth}
		\centering
		\includegraphics[width=.9\linewidth]{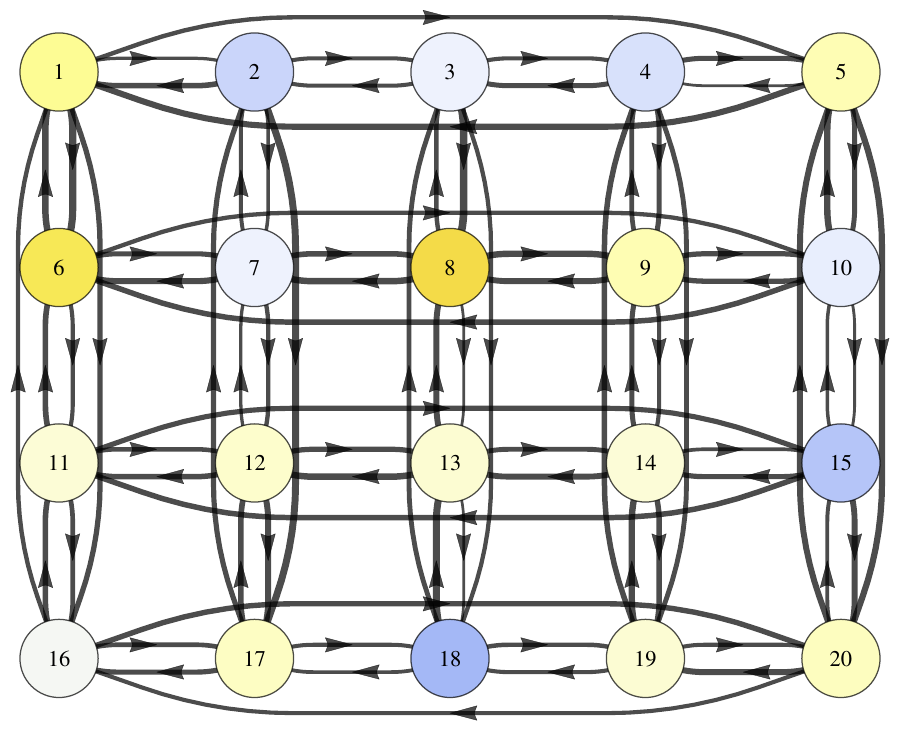}
	\end{subfigure}%
	\begin{subfigure}{.24\textwidth}
		\centering
		\includegraphics[width=.9\linewidth]{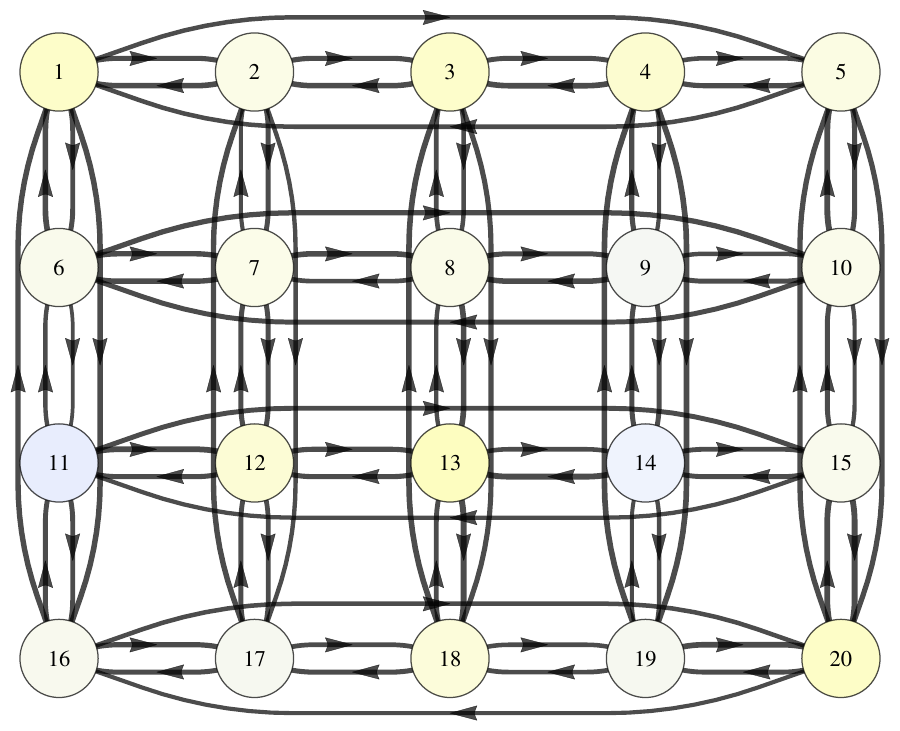}
	\end{subfigure}%
	\begin{subfigure}{.24\textwidth}
		\centering
		\includegraphics[width=.9\linewidth]{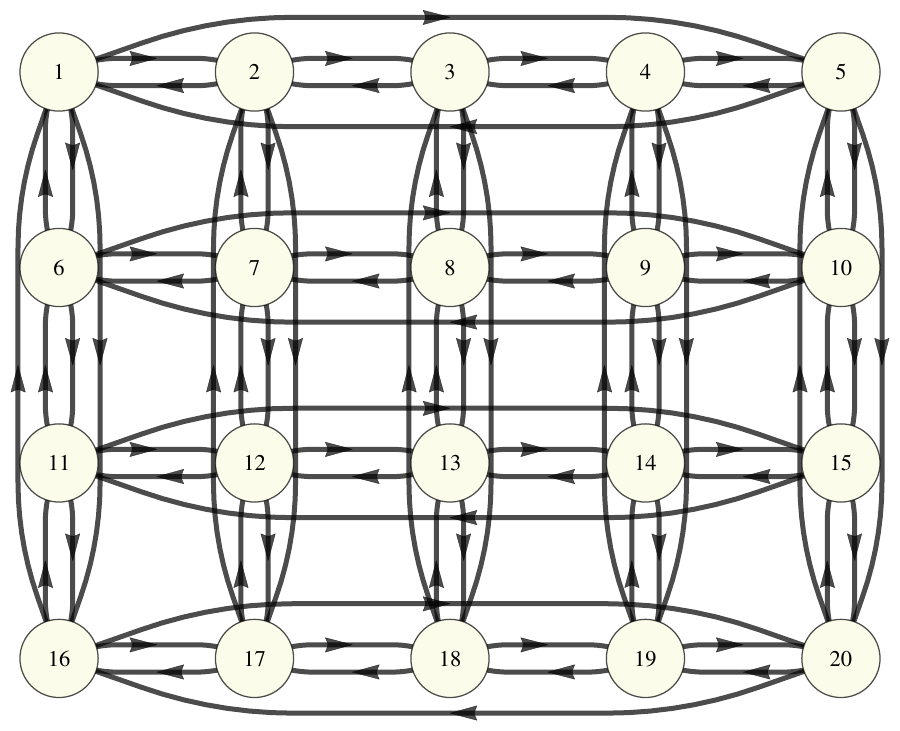}
	\end{subfigure}%
	\\
	\begin{subfigure}{.328\textwidth}
		\raggedright
		\includegraphics[width=.95\linewidth]{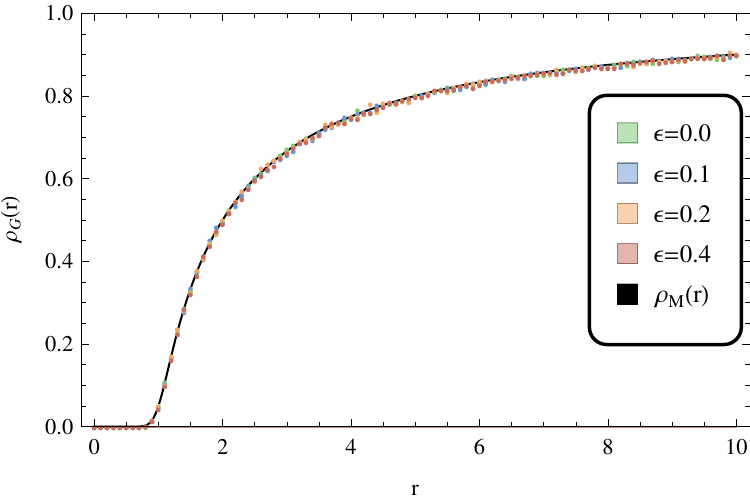}
	\end{subfigure}
	\begin{subfigure}{.328\textwidth}
		\raggedright
		\includegraphics[width=.95\linewidth]{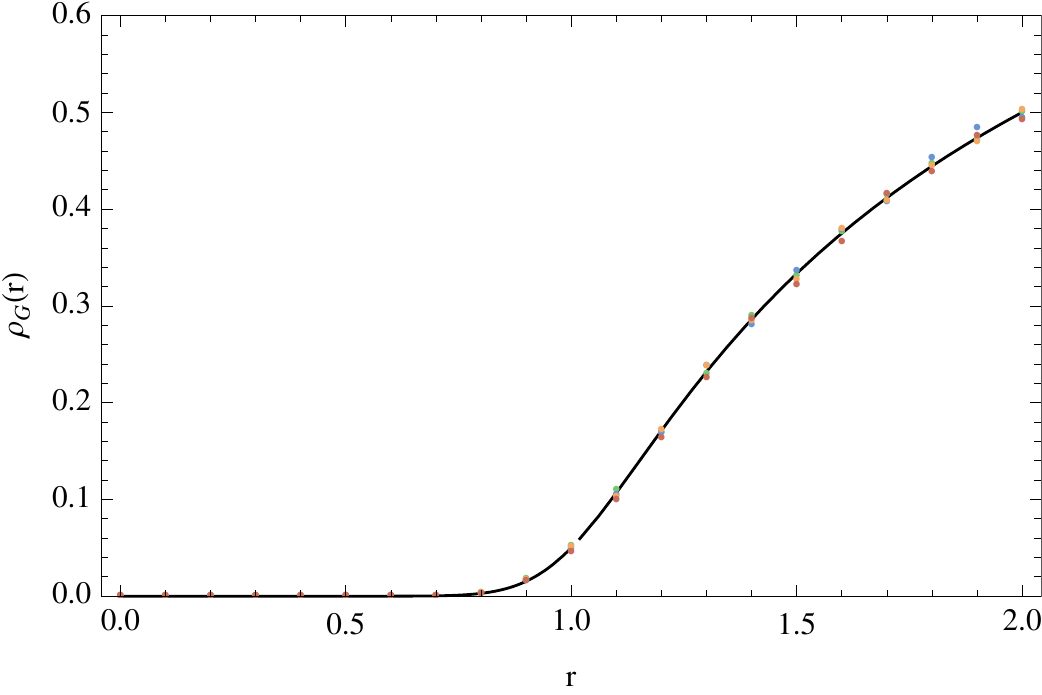}
	\end{subfigure}
	\begin{subfigure}{.328\textwidth}
		\raggedright
		\includegraphics[width=.95\linewidth]{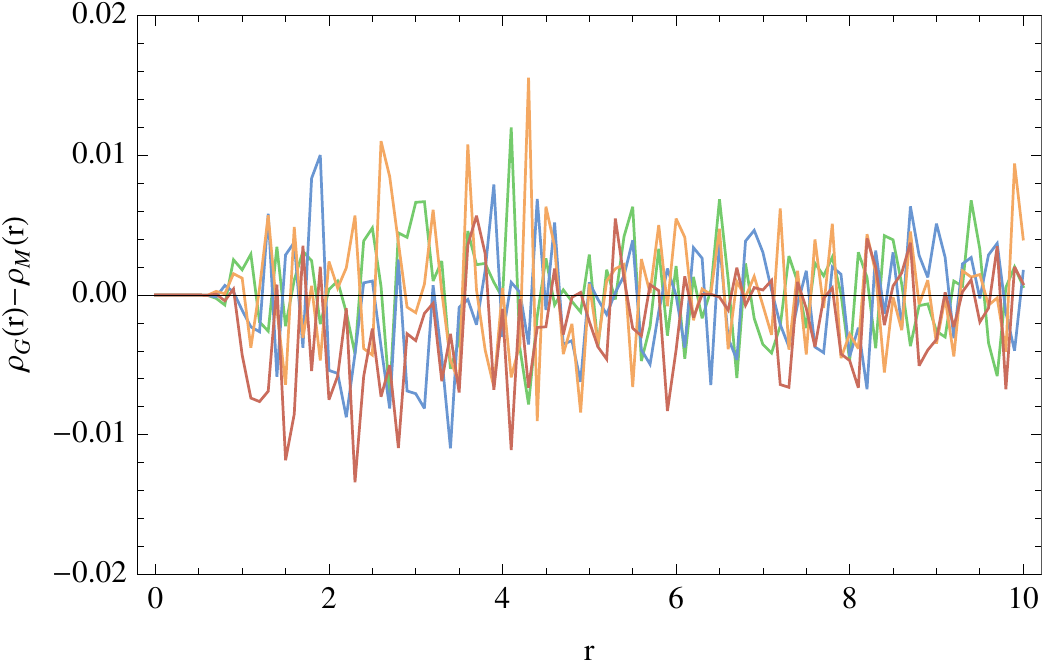}
	\end{subfigure}
	\caption{The robust isothermal theorem guarantees that the fixation probability of an approximately isothermal graph lies in the green region and dashed lines indicate the optimal bound. As the perturbation strength decreases through $\e\in\qa{0.4,0.1}$ and the graph approaches isothermality, the bound improves and converges uniformly to the solid black line, $\rhoM$. The figures of square lattices show how random perturbations shift the graphs from isothermality, as the perturbation strength decreases from left to right; we draw each graph with the directed edges' thickness proportional to their weight and the vertices' color given by the sum of the weights of edges pointing to them. In the bottom row empirical estimates of the fixation probabilities (small circles) are plotted against the values predicted by $\rhoM$ (solid lines) and, despite the perturbations to the graphs, their fixation probabilities lie close to $\rhoM$.}
	\label{fig: robust isothermal theorem}
\end{figure}

This verifies something essential for the process: as in physics, our laws should not depend on arbitrarily small quantities nor make disparate predictions for small perturbations of a system. The {\sc rit} generalizes the isothermal theorem in this sense; if an isothermal graph is perturbed with strength $\e$ such that the assumption \eqref{eq: main assumption} holds, then its fixation probability is close to that of the original graph (Figure \ref{fig: robust isothermal theorem}). There are many ways of rigorously perturbing a graph, so we do not make a precise definition of perturbation here. All we claim is that any perturbation which changes the assumptions of the {\sc rit} continuously can be controlled. The {\sc rit} has many useful applications and is our first ingredient to universality.

Robustness is essential for the analysis of random graphs. We say a random graph model exhibits universal Moran-type behavior if its fixation probability behaves like $\rhoM(r)$ as the graph becomes large. That is, as the graphs become large their macroscopic properties, fixation probabilities, are independent of their microscopic structures, the distributions of individual edges. Mathematically, we ask that the random variable $\sup_{r>0}\absa{\rho_{G_{n}}(r) - \rhoM(r)}$ converges in probability to 0, as $n$ goes to infinity. For finite values of $n$, we can require finer control over this convergence such that
\eqs{\tag{4}
	\P\qa{\sup_{r>0}\absa{\rho_{G_{n}}(r) - \rho_{M}(n,r)} \leq \delta(n) }=1-\e(n),
}
where the functions $\delta(n)=o(1)$ and $\e(n)=o(1)$ can be specified. For the generalized Erd\H{o}s-R\'{e}nyi model \cite{Erdos1960} where edges are produced independently with fixed probability $p$ (see Definitions \ref{def: random graph} and \ref{def: random graph 2}) we prove universality. In Sections \ref{sec: random graphs} and \ref{sec: random graphs 2} we analyze the typical behavior of random graphs and show that with very high probability they satisfy the assumptions of the {\sc rit}, giving us the paper's main result:

\begin{theorem*}
	Let $\pa{G_{n}}_{n\geq 1}$ be a family of random graphs where the directed edge weights are chosen independently according to some suitable distribution (the outgoing edges may be normalized to sum to 1 or not). Then there are constants $C>0$ and $c>0$, not dependent on $n$, such that the fixation probability of a randomly placed mutant of fitness $r>0$ satisfies
	\eq{\tag{5}
		\absa{\rho_{G_{n}}(r)-\rhoM(r)}\leq  \frac{C\pa{\log n}^{C+C\xi}}{\sqrt{n}}
	}
	uniformly in $r$ with probability greater than $1- \exp\p{-\nu \pa{\log n}^{1+\xi}}$, for some positive constants $\xi$ and $\nu$.
\end{theorem*}

This theorem isolates the typical behavior of the Moran process on these random structures. It can be interpreted as stating that random processes generating population structures where vertices and edges are treated independently and interchangeably will almost always produce graphs with Moran-type behavior. While such processes can generate graphs which do not have Moran-type behavior (for example one-rooted or disconnected graphs), these graphs are generated with very low probability as the size of the graphs becomes large. Moreover, it improves upon diffusion approximation methods by explicitly controlling the error rates \cite{Hadjichrysanthou2012}.

The result holds with high probability but sometimes this probability becomes close to 1 only as the graphs become large. The necessary graph size depends on the distribution that the random graph's edge weights are drawn from. In particular, it depends inversely on the parameter $p$ from the generalized Erd\H{o}s-R\'{e}nyi model, which is the probability that there is an edge of some weight between two directed vertices. The smaller this parameter the more disordered and sparse the random graphs and the less uniform their vertices' temperatures, which all tend to decrease the control over the graph's closeness to isothermality, \eqref{eq: main assumption}. Regardless, our choice of the parameter $\pa{\log n}^{1+\xi}$ guarantees that the bound [\ref{eq: in out bound}] decays to 0 and that it holds with probability approaching 1 as $n$ becomes large.

\begin{figure}[H]
	\centering
	\begin{subfigure}{.02\textwidth}
		\raggedright
		\includegraphics[width=1\linewidth]{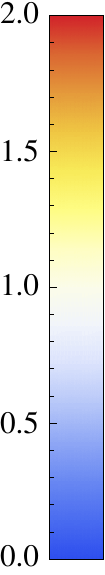}
	\end{subfigure}%
	\begin{subfigure}{.12\textwidth}
		\raggedleft
		\includegraphics[width=.9\linewidth]{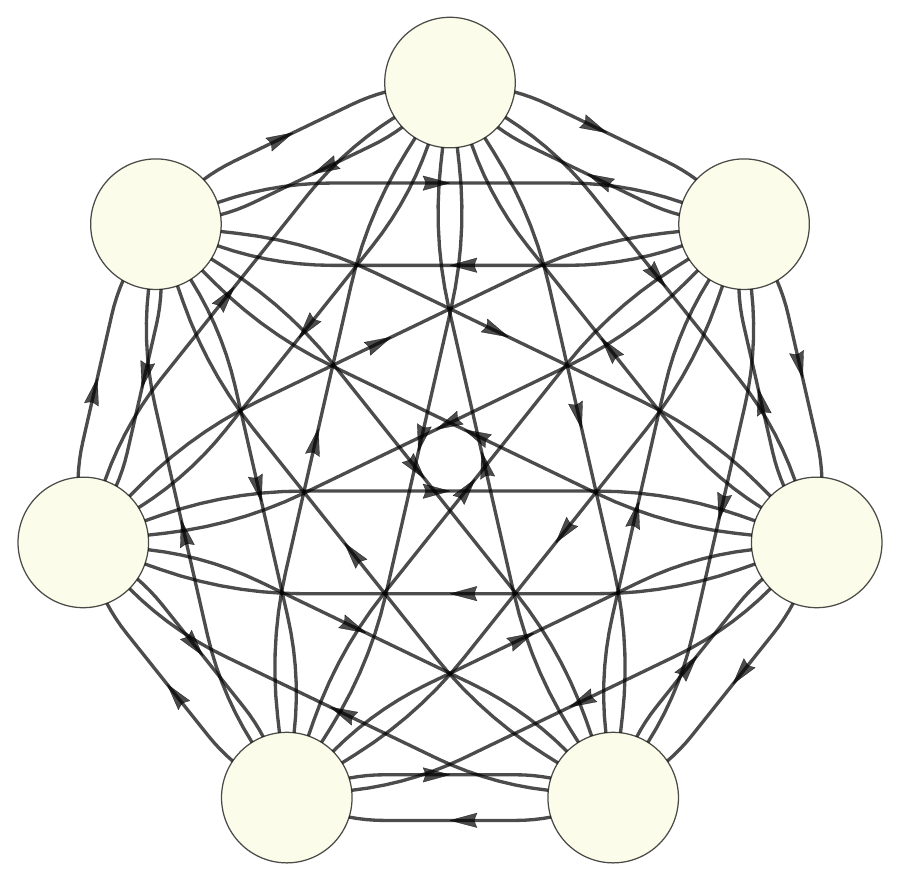}
	\end{subfigure}%
	\begin{subfigure}{.2\textwidth}
		\raggedright
		\includegraphics[width=.9\linewidth]{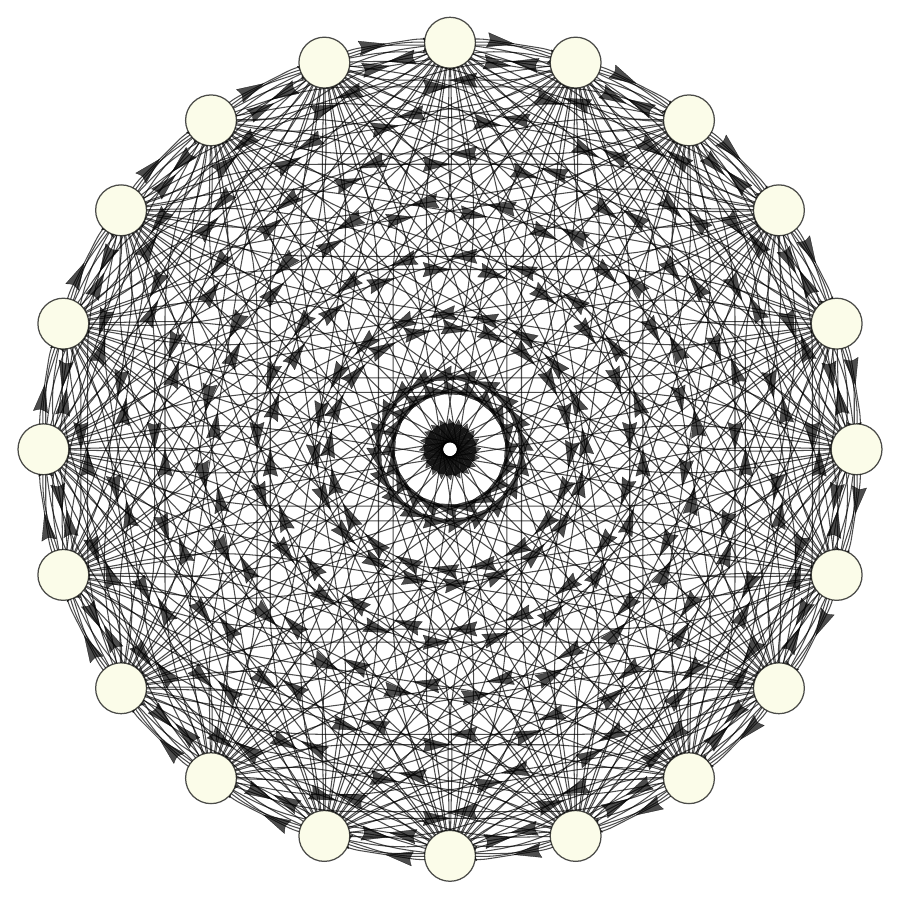}
	\end{subfigure}%
	\begin{subfigure}{.12\textwidth}
		\raggedleft
		\includegraphics[width=.9\linewidth]{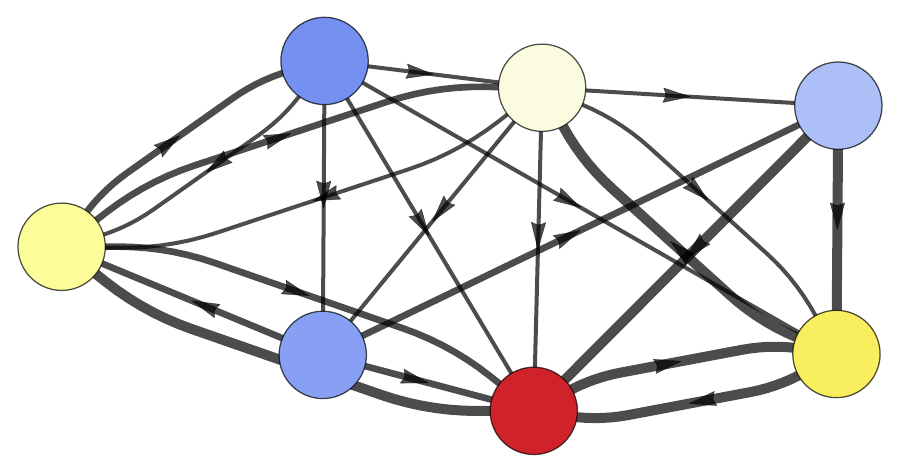}
	\end{subfigure}%
	\begin{subfigure}{.2\textwidth}
		\raggedright
		\includegraphics[width=.9\linewidth]{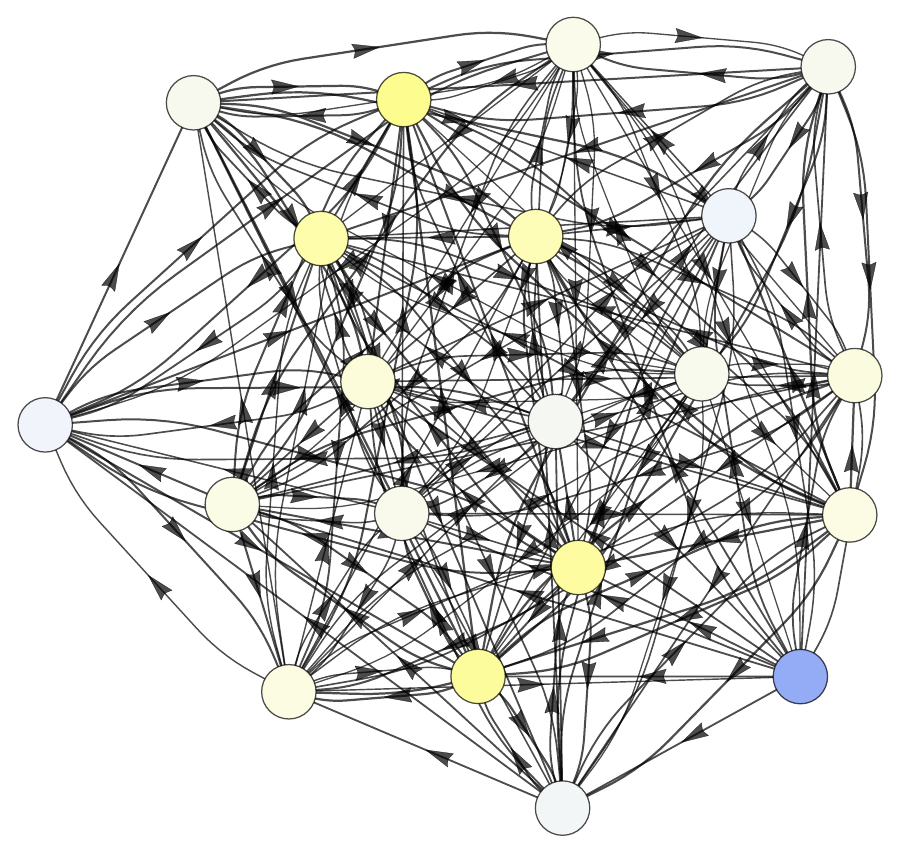}
	\end{subfigure}%
	\begin{subfigure}{.12\textwidth}
		\raggedleft
		\includegraphics[width=.9\linewidth]{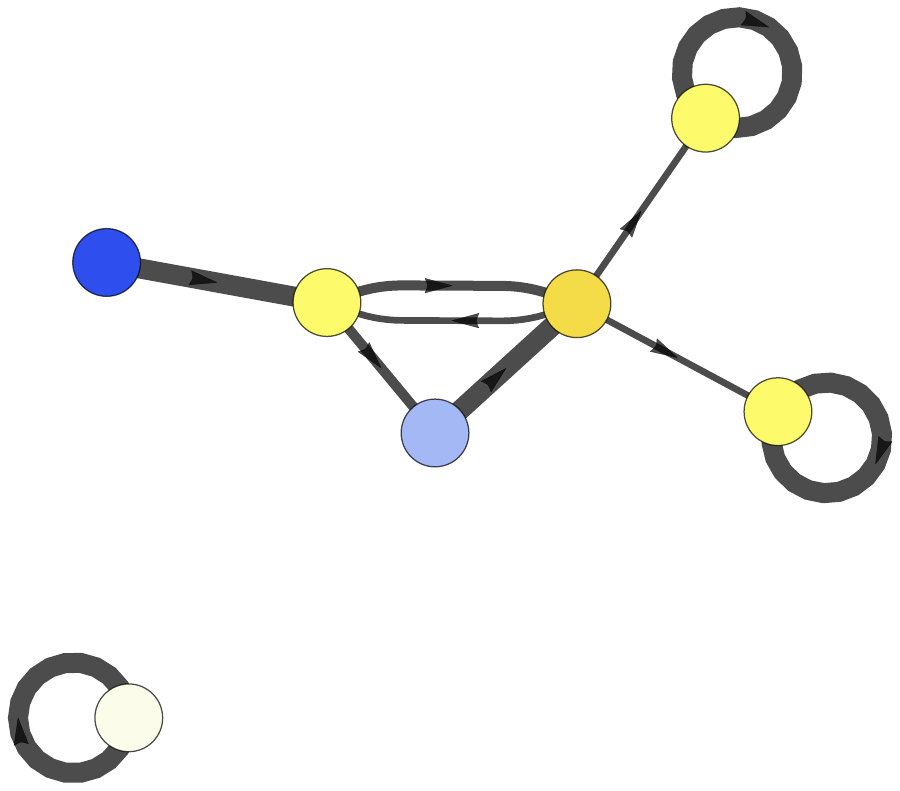}
	\end{subfigure}
	\begin{subfigure}{.2\textwidth}
		\raggedright
		\includegraphics[width=.9\linewidth]{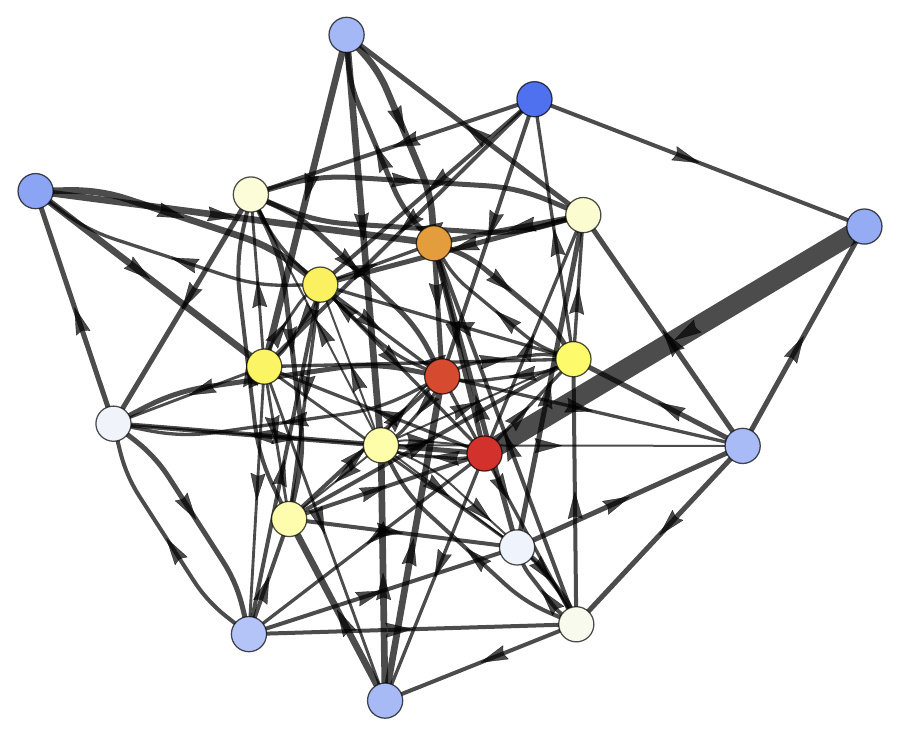}
	\end{subfigure}
	\\
	\begin{subfigure}{.33\textwidth}
		\raggedright
		\includegraphics[width=.95\linewidth]{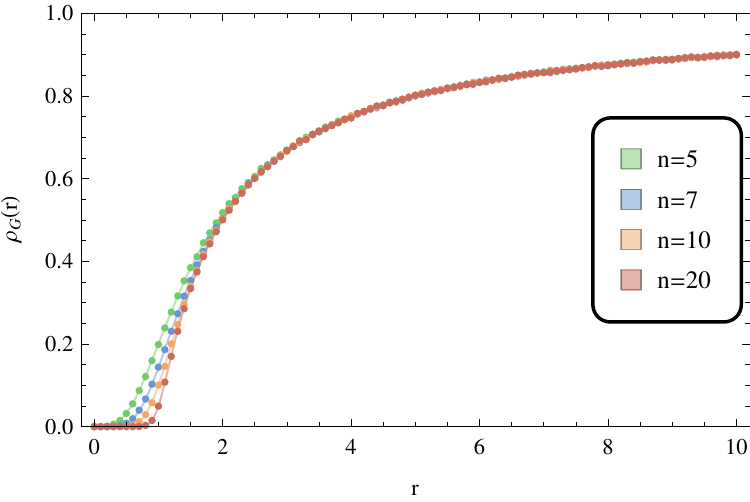}
	\end{subfigure}%
	\begin{subfigure}{.33\textwidth}
		\raggedright
		\includegraphics[width=.95\linewidth]{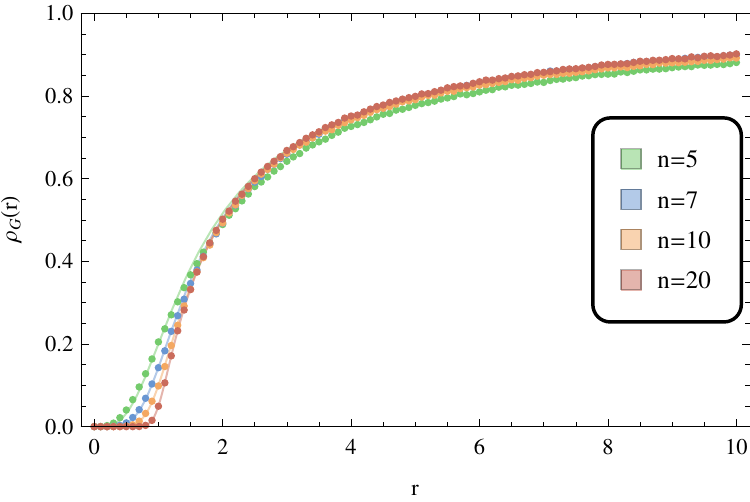}
	\end{subfigure}%
	\begin{subfigure}{.33\textwidth}
		\raggedright
		\includegraphics[width=.95\linewidth]{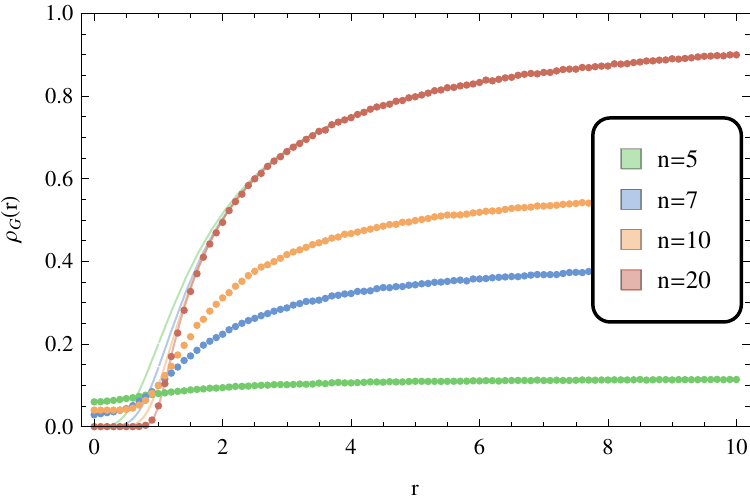}
	\end{subfigure}
	\\
	\begin{subfigure}{.33\textwidth}
		\raggedright
		\includegraphics[width=.95\linewidth]{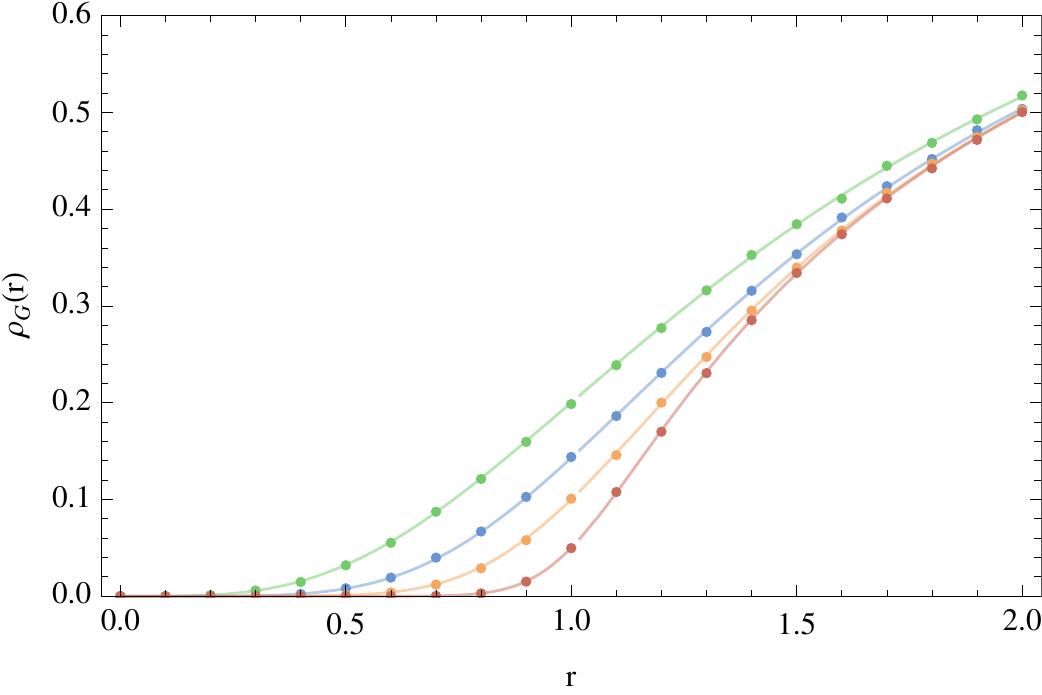}
	\end{subfigure}%
	\begin{subfigure}{.33\textwidth}
		\raggedright
		\includegraphics[width=.95\linewidth]{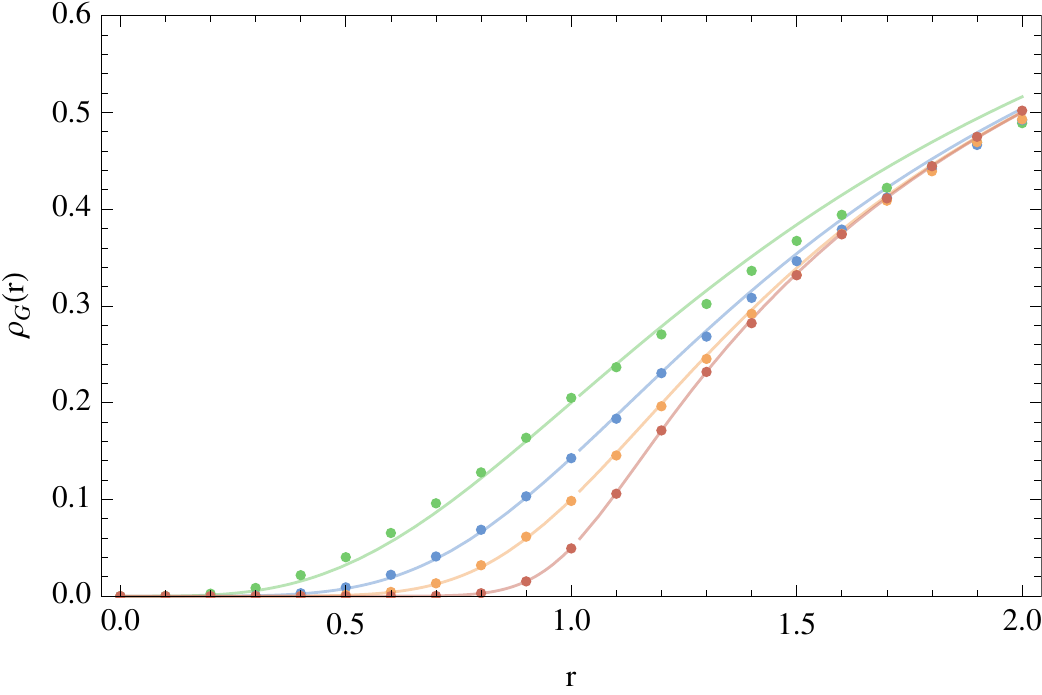}
	\end{subfigure}%
	\begin{subfigure}{.33\textwidth}
		\raggedright
		\includegraphics[width=.95\linewidth]{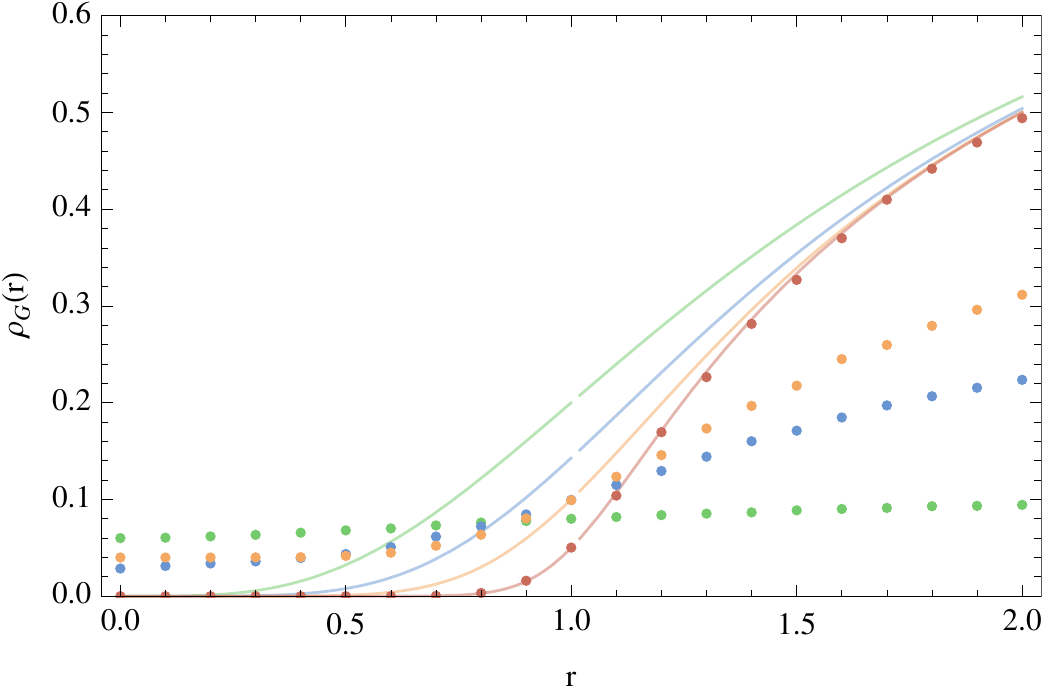}
	\end{subfigure}
	\\
	\begin{subfigure}{.33\textwidth}
		\raggedright
		\includegraphics[width=.95\linewidth]{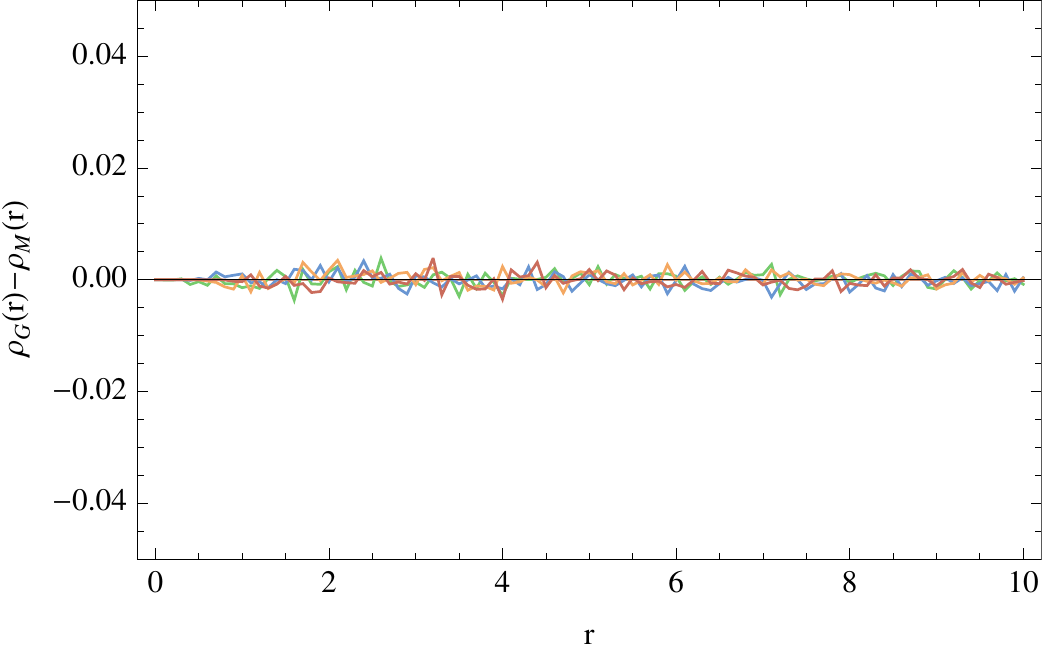}
	\end{subfigure}%
	\begin{subfigure}{.33\textwidth}
		\raggedright
		\includegraphics[width=.95\linewidth]{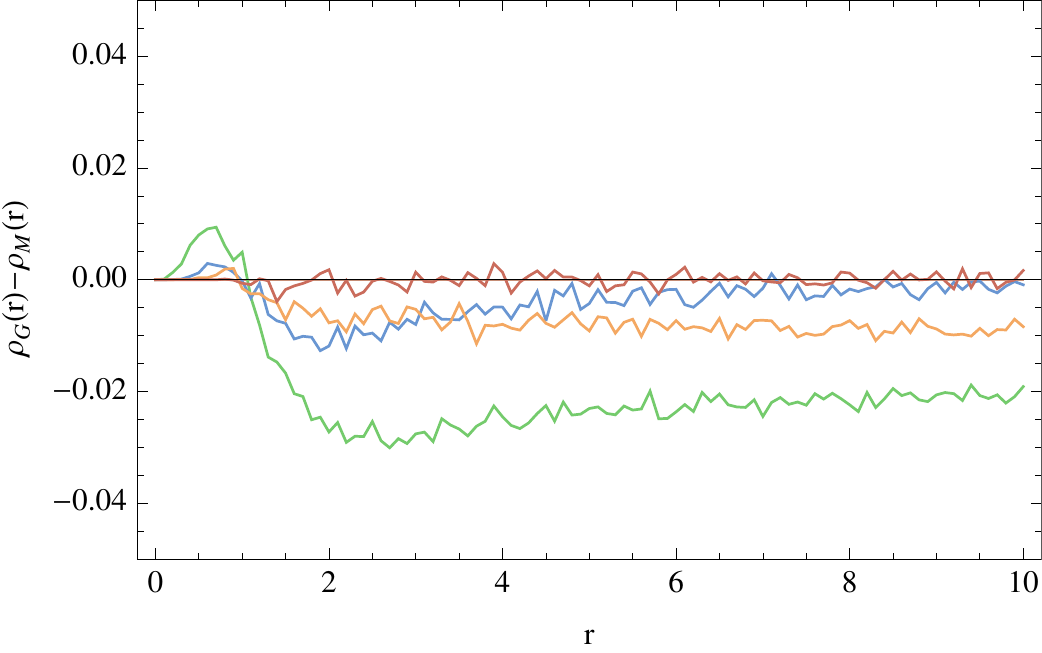}
	\end{subfigure}%
	\begin{subfigure}{.33\textwidth}
		\raggedright
		\includegraphics[width=.95\linewidth]{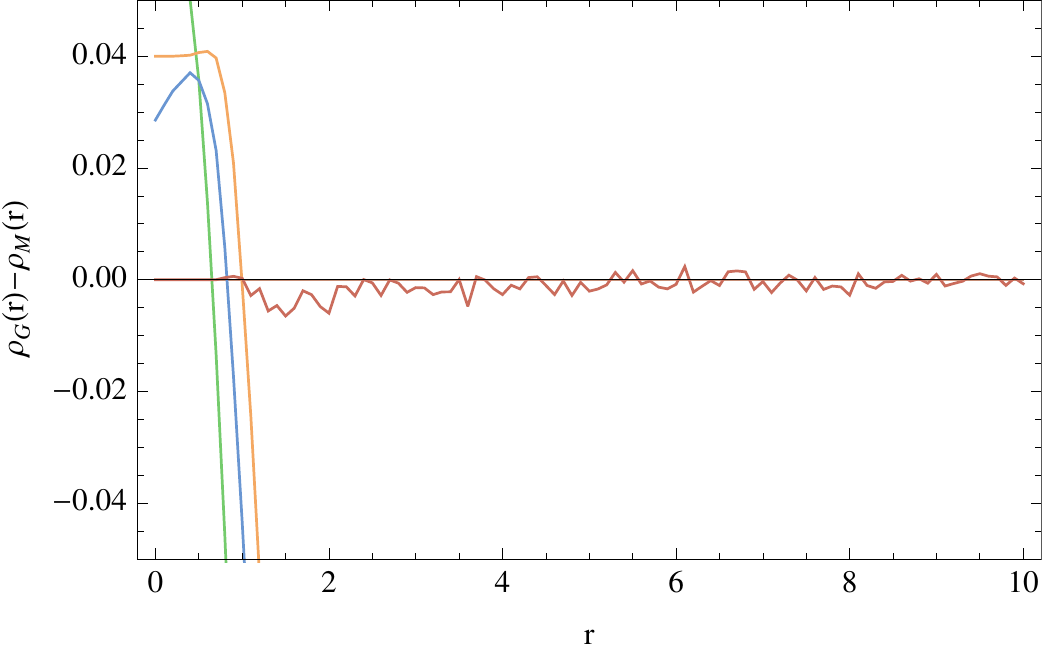}
	\end{subfigure}
	\caption{The fixation probability of the generalized Erd\H{o}s-R\'{e}nyi random graphs converge uniformly to $\rhoM$. The three columns from left to right correspond to Erd\H{o}s-R\'{e}nyi random graphs with decreasing connection probabilities $p=1$, $p=0.6$, and $p=0.3$. The representative random graphs in the top row show both the increasing sparsity and disorder as $p$ decreases and the elimination of degeneracy (rootedness and disconnectedness) and the increasing uniformity of temperature as the graph sizes increase. In the middle two rows empirical estimates of the fixation probabilities (small circles) are plotted against the values predicted by $\rhoM$ (solid lines). When $p=1$ the graphs are isothermal and thus correspond exactly to their predicted values which can be seen even more clearly in the bottom row, where the difference of the empirical fixation probabilities and their predicted values display as stochastic fluctuations about 0. For $p=0.6$ and $p=0.3$, the convergence of the empirical values to $\rhoM$ as the graphs increases in size is apparent. Smaller graphs are typically repressors as illustrated by the clear sign change at $r=1$ in the difference of empirical and predicted values, whereas larger graphs fluctuate about 0. Moreover, the convergence is patently slower in $n$ for smaller values of $p$.}
	\label{fig: random graphs}
\end{figure}

We investigated the issues of convergence for small values of $n$ numerically to illustrate our analytical result (Figure \ref{fig: random graphs}). For Erd\H{o}s-R\'{e}nyi random graphs (see Section \ref{sec: random graphs} with the distribution chosen as Bernoulli), we generated 10 random graphs according to the procedure outlined in Definition \ref{def: random graph} for fixed values of $0<p<1$. On each graph the Moran process was simulated $10^{4}$ times for various values of $0\leq r\leq 10$ to give the empirical fixation probability, that is, the proportion of times that the mutant fixed in the simulation. Degenerate graphs were not excluded from the simulations but rather than estimating their fixation probabilities, we calculated them exactly, so that 1-rooted graphs were given fixation probability $1/n$ and many-rooted and disconnected graphs were given fixation probability $0$. Trivially, such 1-rooted graphs are repressors---that is, the fixation probability of a mutant of fitness $0<r<1$ (and a mutant of fitness $r>1$) is greater than (and less than respectively) the mutant's fixation probability in a well-mixed population---but repressor graphs without these degenerate properties were also observed. As the graphs become larger their fixation probabilities match $\rhoM$ closely and degeneracy becomes highly improbable as predicted by our result.

\begin{figure}[H]
	\centering
	\begin{subfigure}{.02\textwidth}
		\raggedright
		\includegraphics[width=1\linewidth]{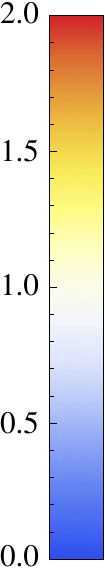}
	\end{subfigure}%
	\begin{subfigure}{.2\textwidth}
		\centering
		\includegraphics[width=.8\linewidth]{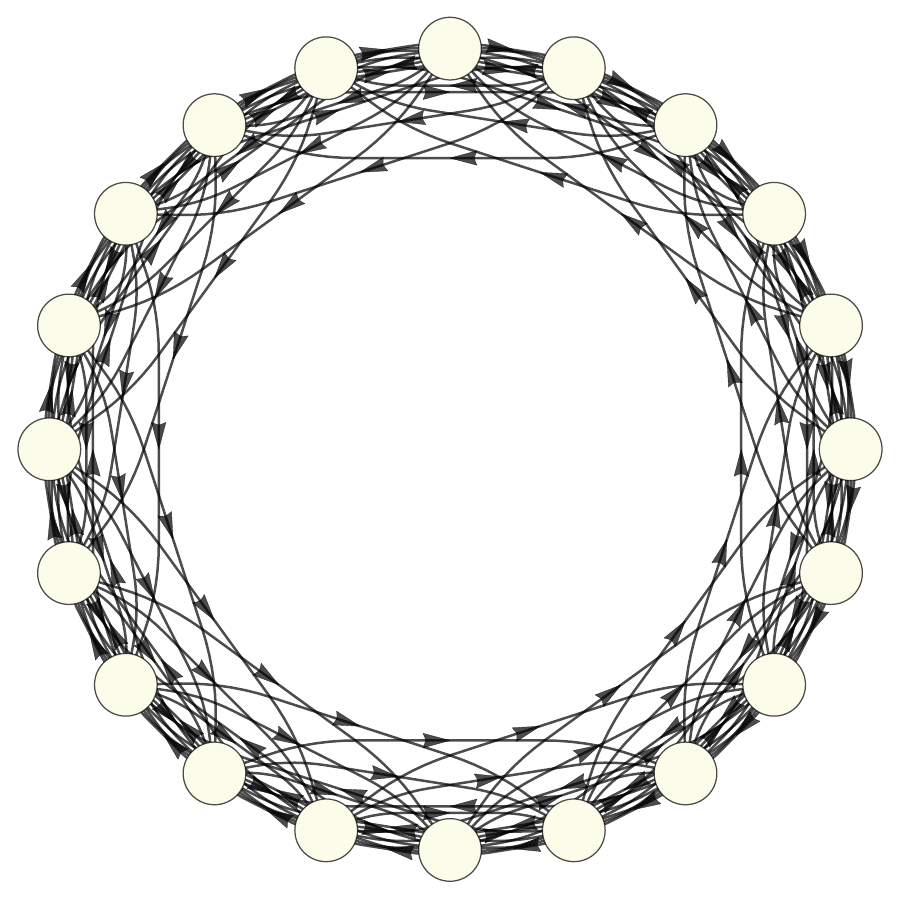}
	\end{subfigure}%
	\begin{subfigure}{.2\textwidth}
		\centering
		\includegraphics[width=.8\linewidth]{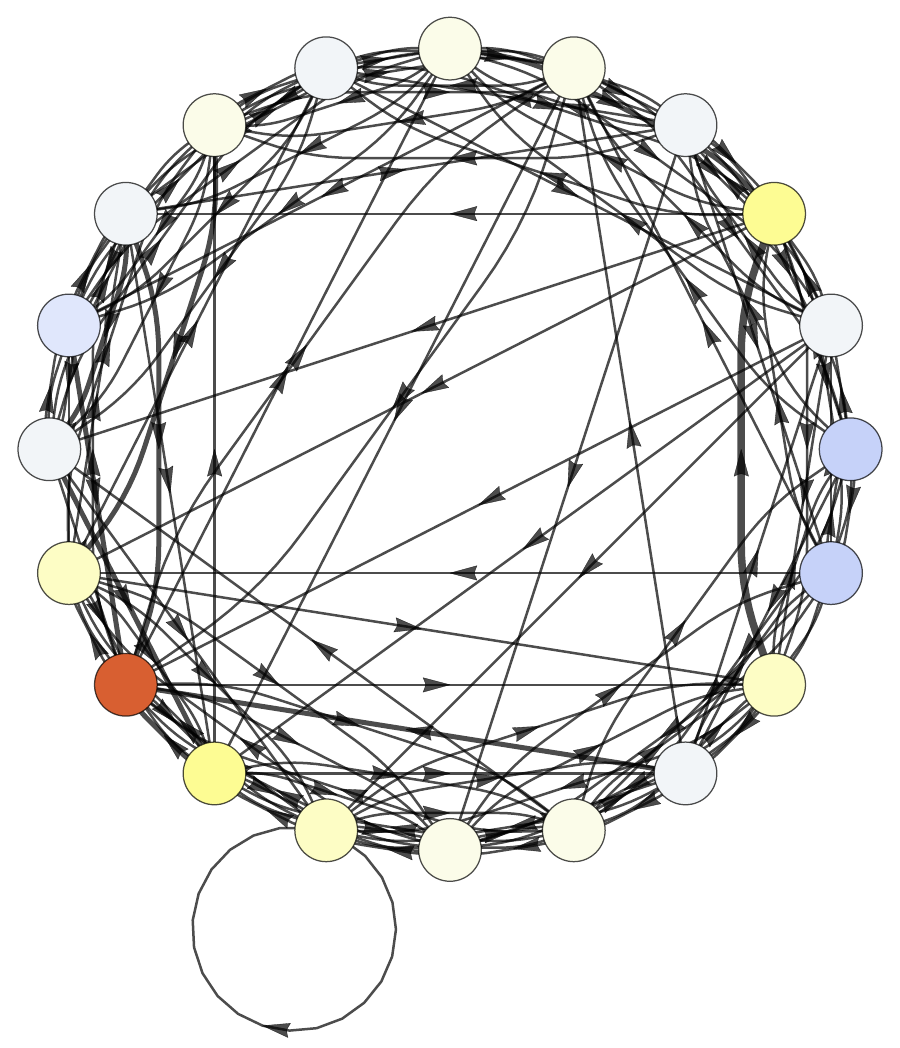}
	\end{subfigure}%
	\begin{subfigure}{.2\textwidth}
		\centering
		\includegraphics[width=.9\linewidth]{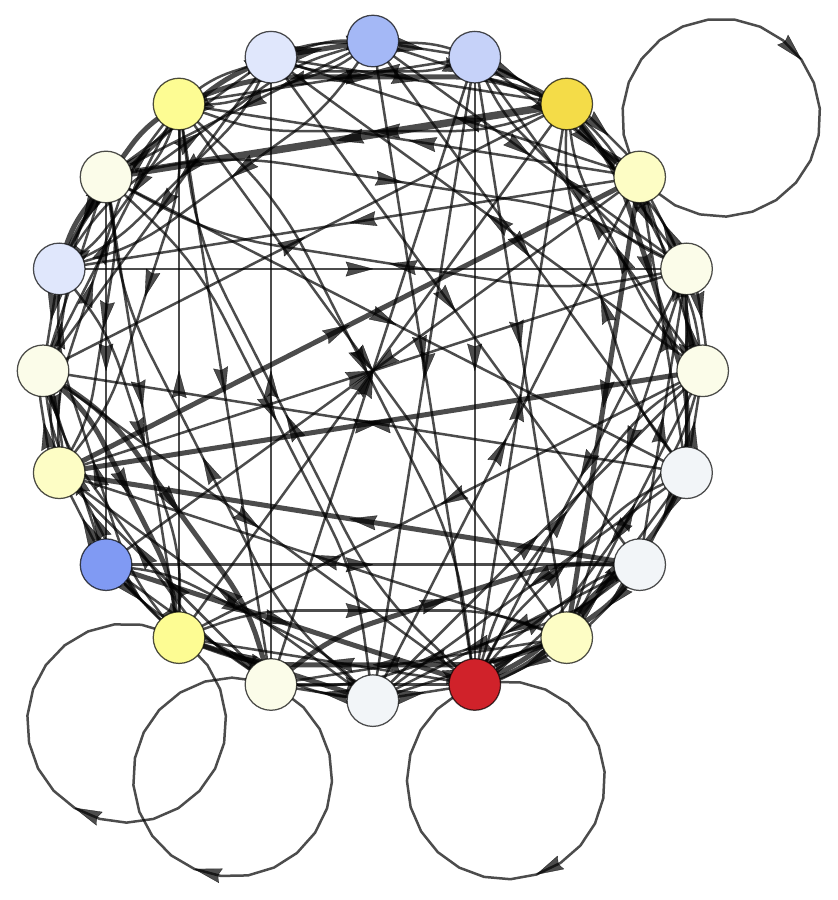}
	\end{subfigure}%
	\begin{subfigure}{.2\textwidth}
		\centering
		\includegraphics[width=1.2\linewidth]{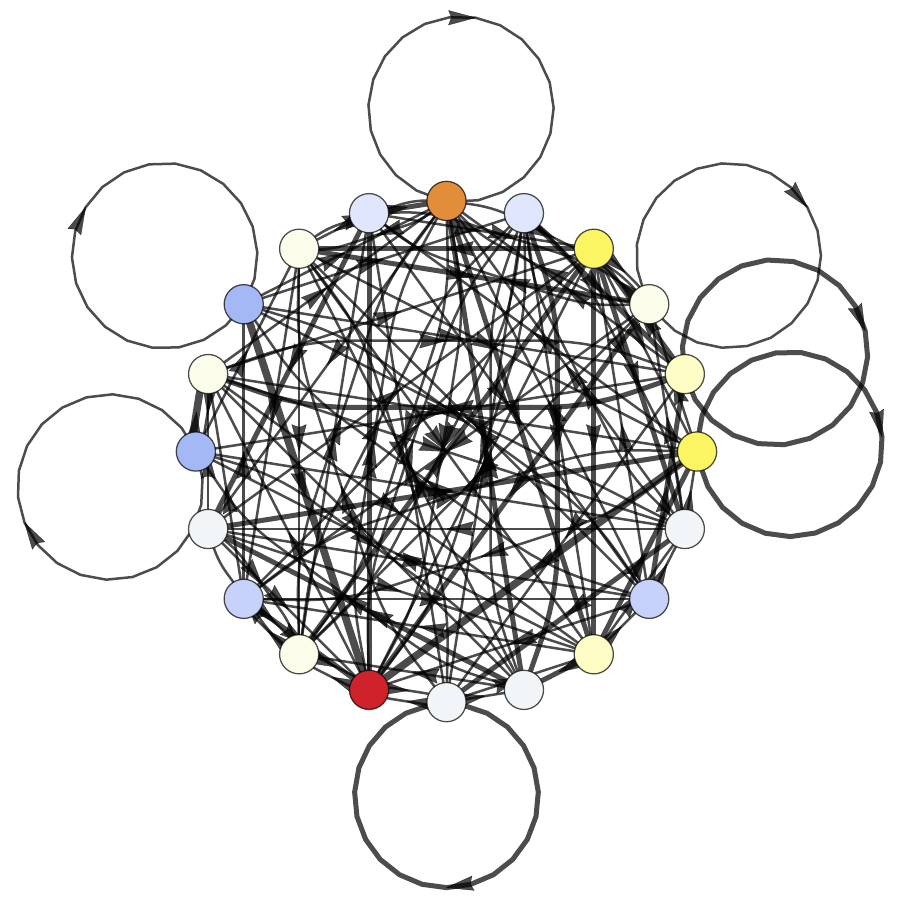}
	\end{subfigure}%
	\\
	\begin{subfigure}{.328\textwidth}
		\raggedright
		\includegraphics[width=.95\linewidth]{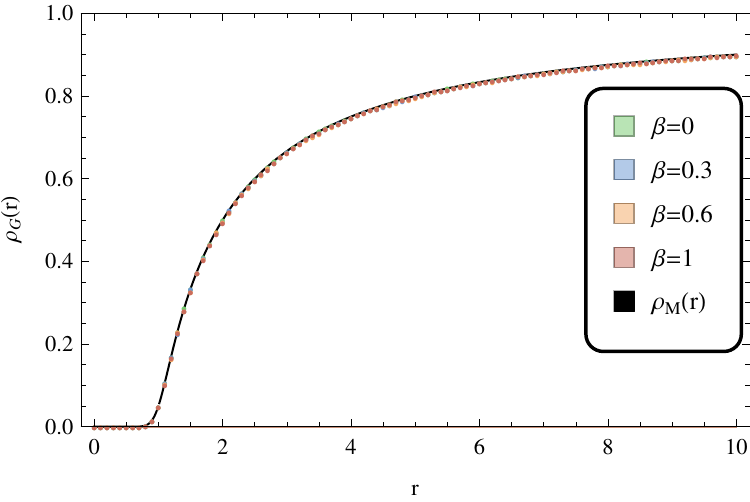}
	\end{subfigure}
	\begin{subfigure}{.328\textwidth}
		\raggedright
		\includegraphics[width=.95\linewidth]{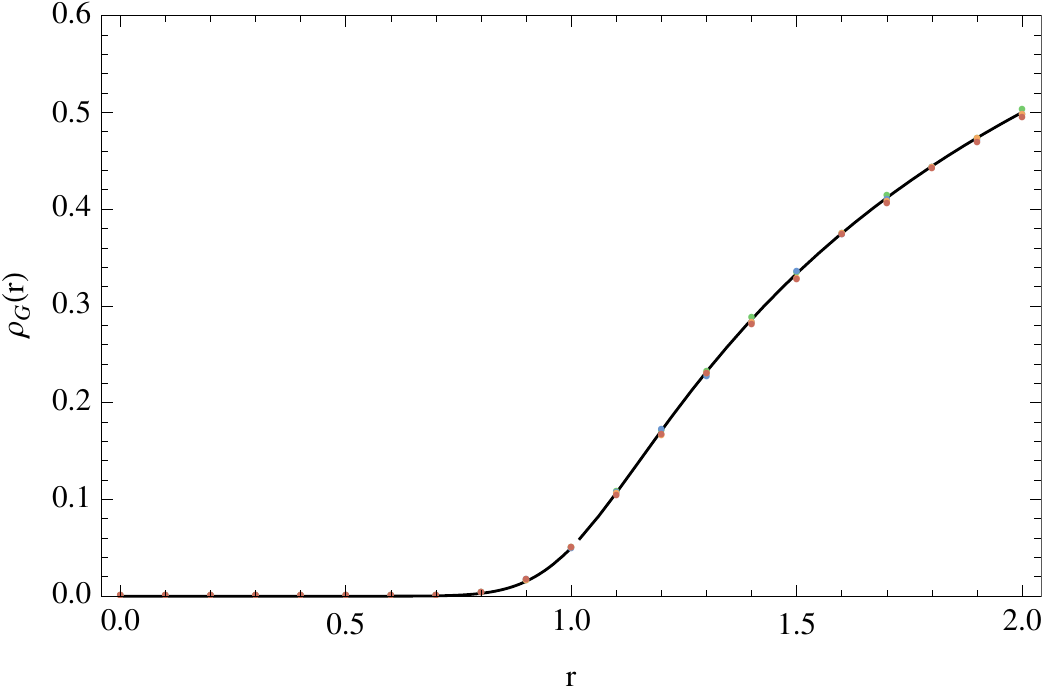}
	\end{subfigure}
	\begin{subfigure}{.328\textwidth}
		\raggedright
		\includegraphics[width=.95\linewidth]{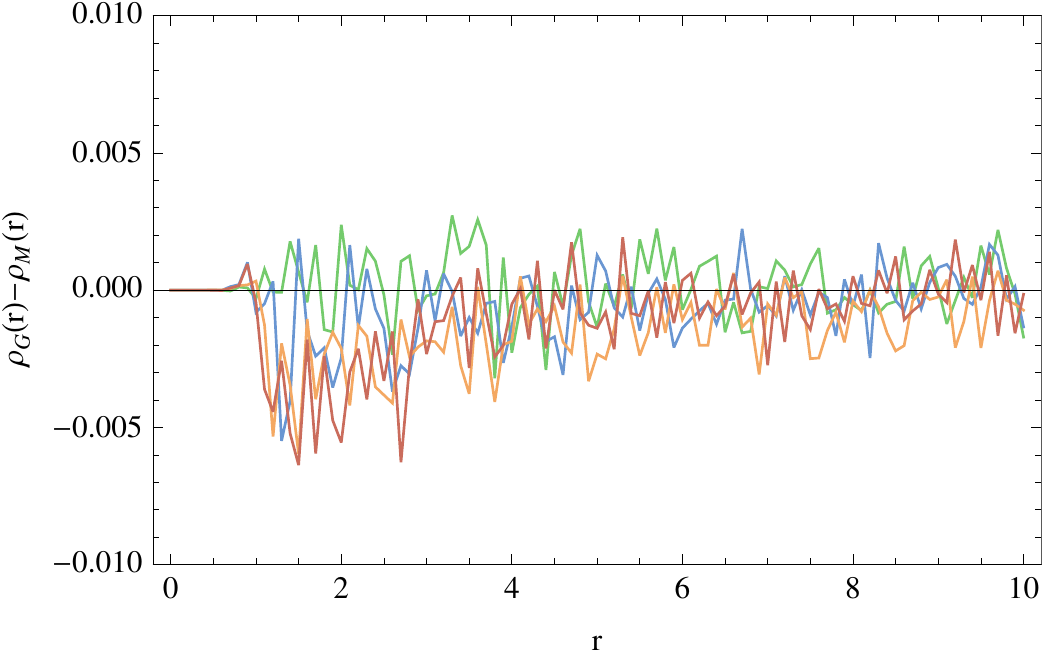}
	\end{subfigure}
	\caption{Small-world networks also show universal behavior. Representative Watts-Strogatz random graphs display increasing disorder as the rewiring probability $\beta$ increases from $0$ to $1$, which may be viewed as an interpolation between an isothermal graph to an Erd\H{o}s-R\'{e}nyi random graph. For all values of $\beta$ the correspondence to $\rhoM$ is striking but mathematical proof is lacking.}
	\label{fig: ws}
\end{figure}

In addition to the generalized Erd\H{o}s-R\'{e}nyi random graphs, we also considered the Watts-Strogatz model and the Barab\'{a}si-Albert model. The Watts-Strogatz model \cite{Watts1998} produces random graphs with small-world properties, that is, high clustering and short average path length. The model has three inputs: a parameter $0\leq\beta\leq1$, the graph size $n$, and the mean degree $2k$. Typically, the model produces random, undirected graphs, thus, to escape isothermality, it was modified slightly to produce weighted, directed graphs. We do this in the most natural way: we start with a directed $2k$-regular graph where each node is connected to its $2k$ nearest neighbors if the graph is arranged on a cycle (see Figure \ref{fig: ws}), and then we rewire each edge to a new vertex chosen uniformly at random with probability $\beta$ independently. Since the number of edges leaving each vertex is fixed at $2k$, the weight of each edge is exactly $1/(2k)$. Potentially, there can be multiple edges for one vertex to another, which we account for by summing the edge weights. The model may be viewed as an interpolation between an isothermal, $2k$-regular graph and an Erd\H{o}s-R\'{e}nyi graph by the parameter $\beta$.

Moran-type behavior was observed in the Watts-Strogatz model for all values of the input parameters we simulated (Figure \ref{fig: ws}). While mathematical proof of universality in the Watts-Strogatz model is still needed, there is hope that the techniques of this paper may be applied in this situation as the in-degrees of the vertices are concentrated around 1 for graphs with large degree $2k$.

Unlike the Erd\H{o}s-R\'{e}nyi and Watts-Strogatz models, scale-free networks are random graphs where the in-degrees of the vertices follow a power law. Normally, scale-free networks are undirected and unweighted. To produce weighted, directed scale-free networks, we modified the preferential attachment algorithm of Barab\'{a}si-Albert \cite{Barabasi1999}: we start with a connected cycle and then add directed edges of equal weight in sequence to a randomly selected vertex where the destination of each edge is selected proportional to the in-degree of the current vertices. 

\begin{figure}[H]
	\centering
	\begin{subfigure}{.02\textwidth}
		\raggedright
		\includegraphics[width=1\linewidth]{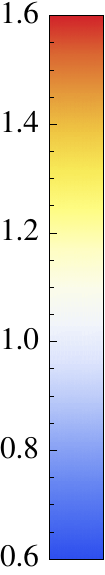}
	\end{subfigure}%
	\begin{subfigure}{.2\textwidth}
		\raggedright
		\includegraphics[width=1\linewidth]{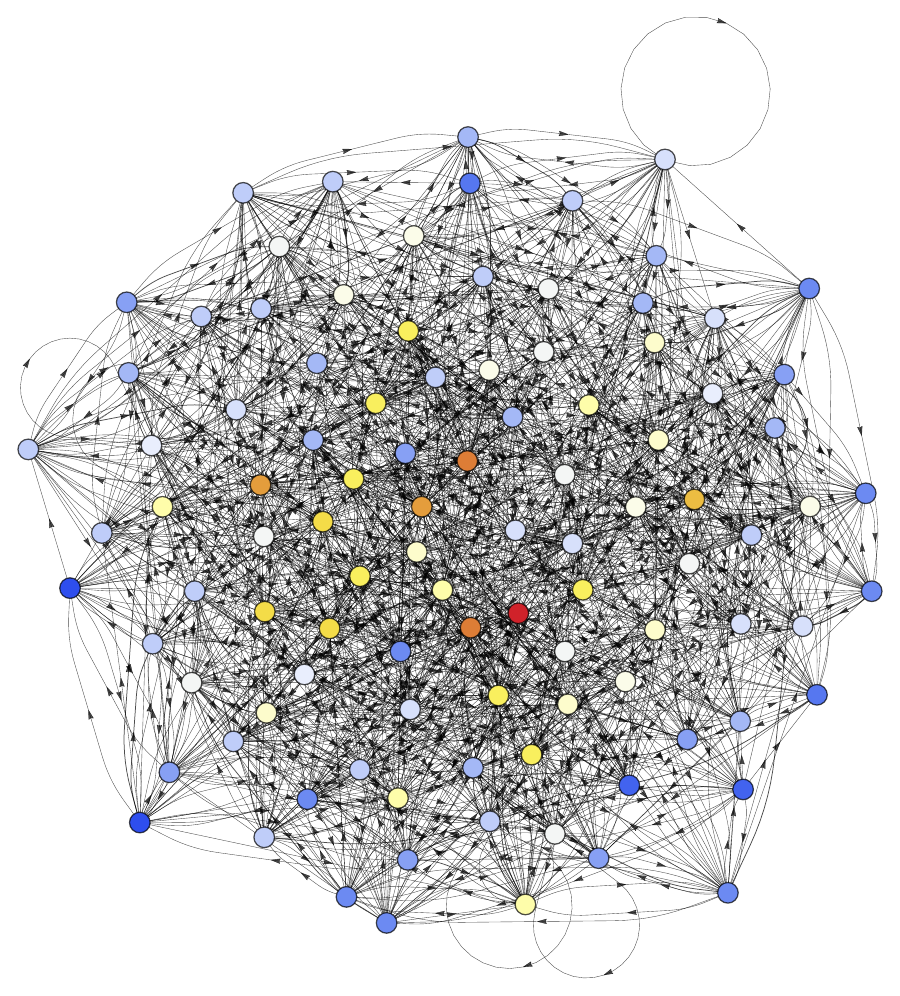}
	\end{subfigure}%
	\begin{subfigure}{.23\textwidth}
		\raggedright
		\includegraphics[width=.95\linewidth]{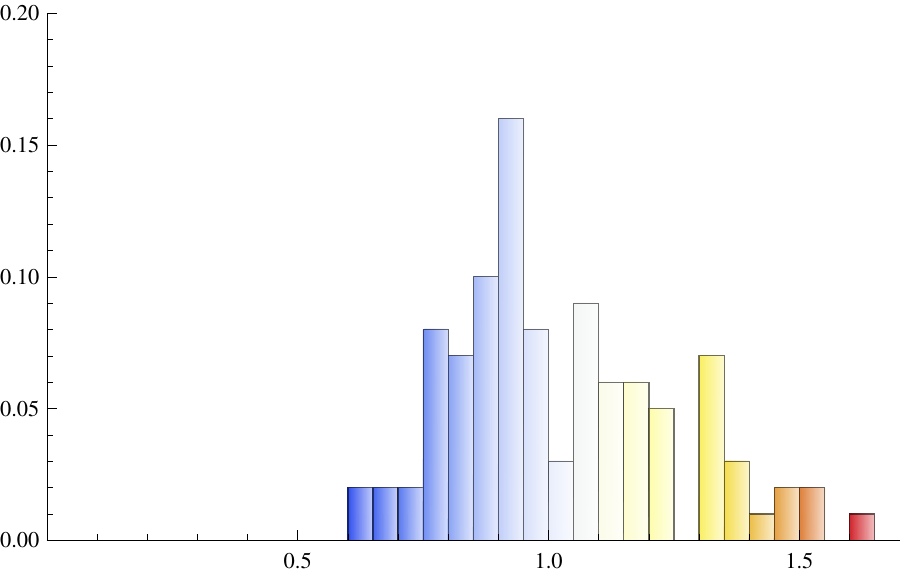}
	\end{subfigure}
	\begin{subfigure}{.02\textwidth}
		\raggedleft
		\includegraphics[width=1\linewidth]{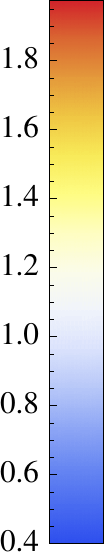}
	\end{subfigure}%
	\begin{subfigure}{.23\textwidth}
		\raggedleft
		\includegraphics[width=1\linewidth]{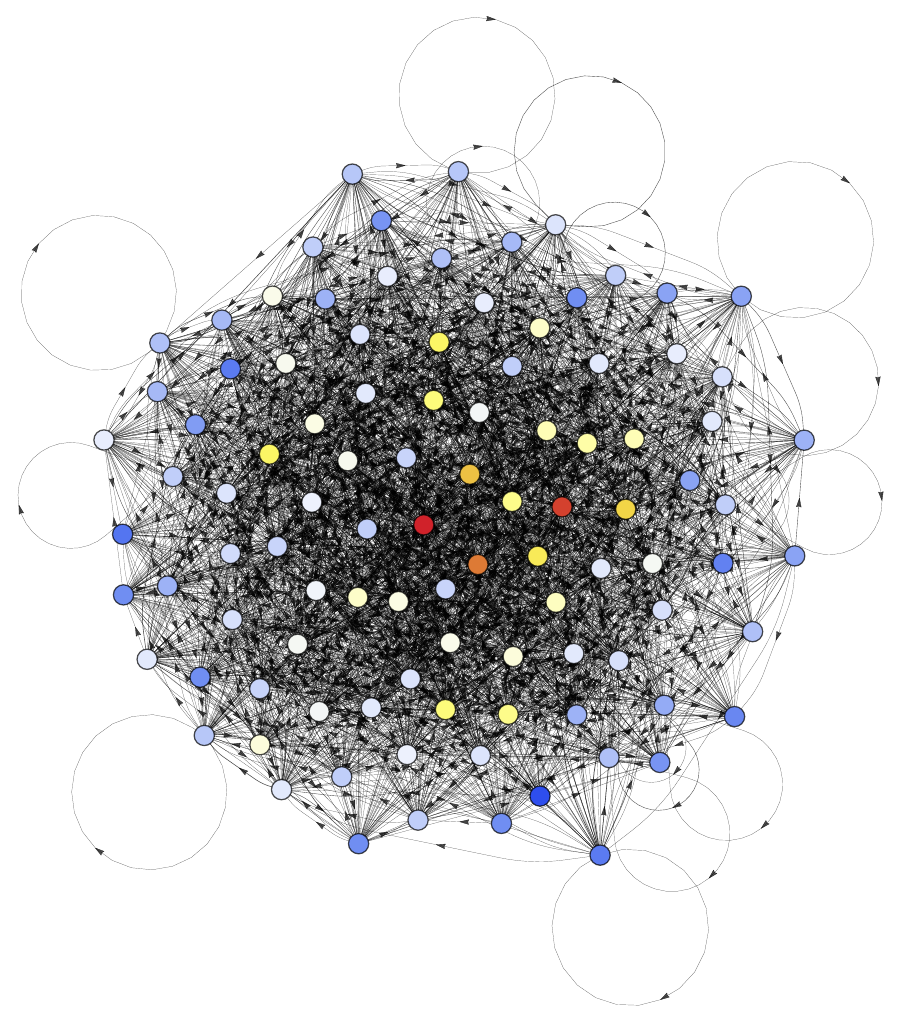}
	\end{subfigure}%
	\begin{subfigure}{.23\textwidth}
		\raggedleft
		\includegraphics[width=.95\linewidth]{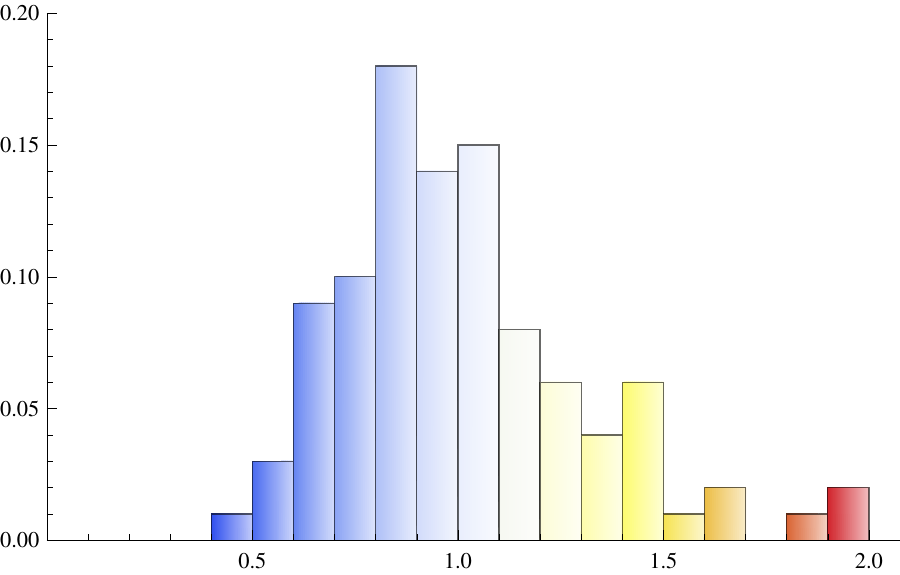}
	\end{subfigure}
	\\
	\begin{subfigure}{.02\textwidth}
		\raggedright
		\includegraphics[width=1\linewidth]{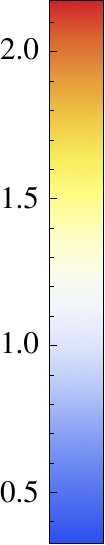}
	\end{subfigure}%
	\begin{subfigure}{.23\textwidth}
		\raggedright
		\includegraphics[width=1\linewidth]{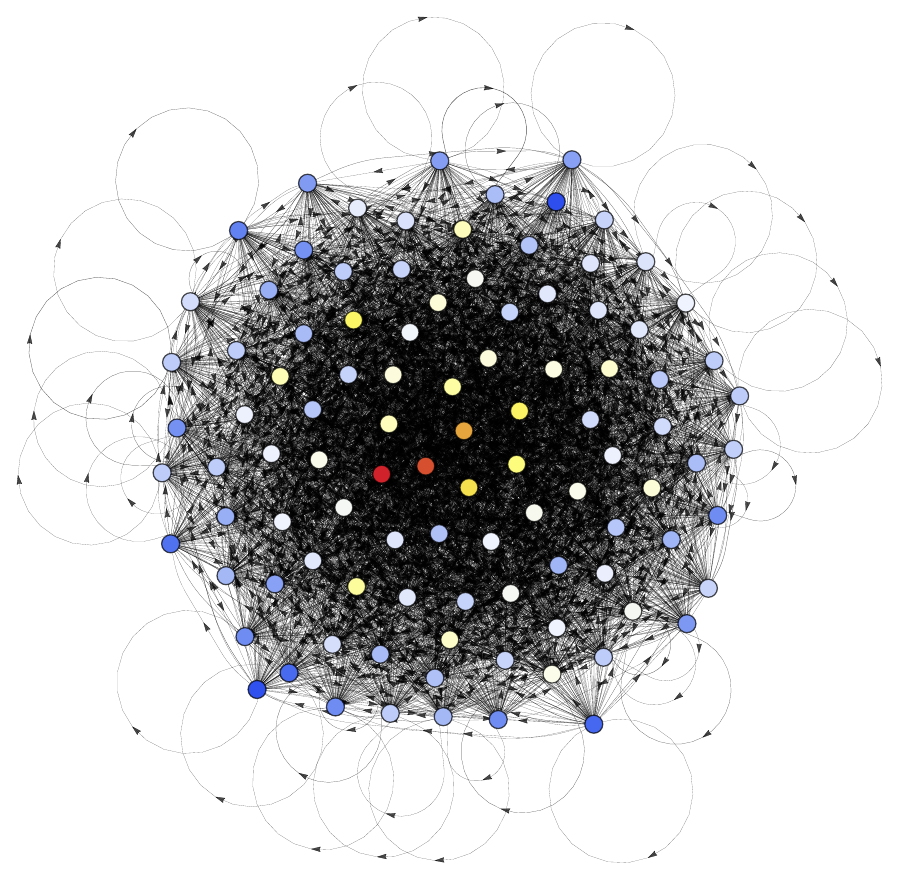}
	\end{subfigure}
	\begin{subfigure}{.23\textwidth}
		\raggedright
		\includegraphics[width=.95\linewidth]{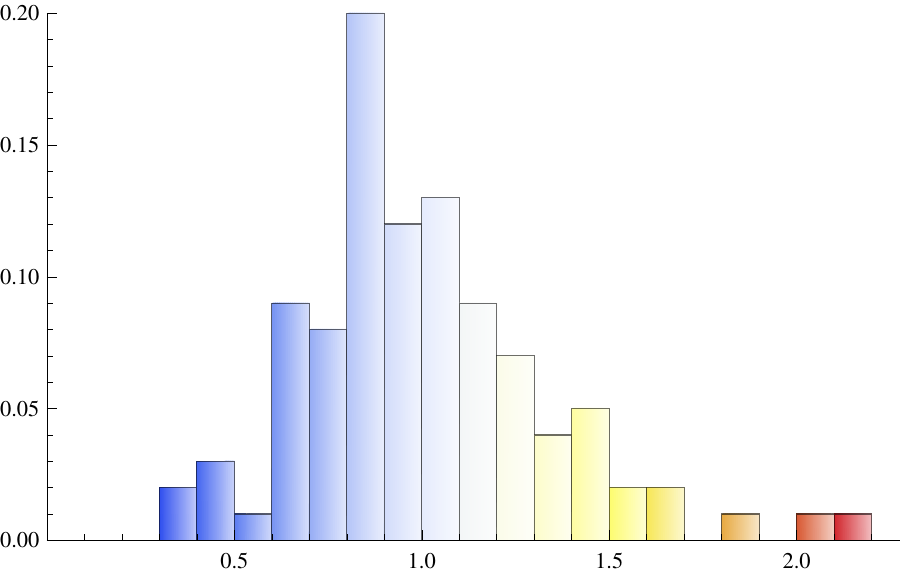}
	\end{subfigure}
	\begin{subfigure}{.02\textwidth}
		\raggedleft
		\includegraphics[width=1\linewidth]{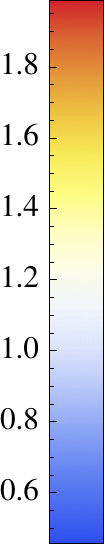}
	\end{subfigure}%
	\begin{subfigure}{.23\textwidth}
		\raggedleft
		\includegraphics[width=1\linewidth]{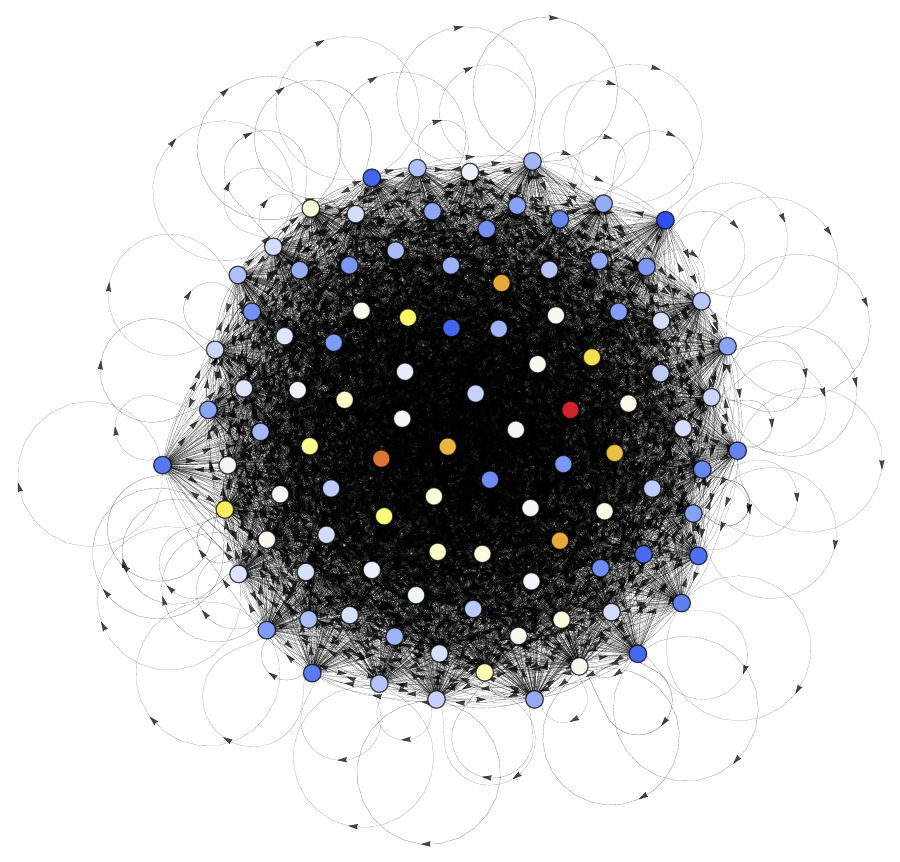}
	\end{subfigure}
	\begin{subfigure}{.23\textwidth}
		\raggedleft
		\includegraphics[width=.95\linewidth]{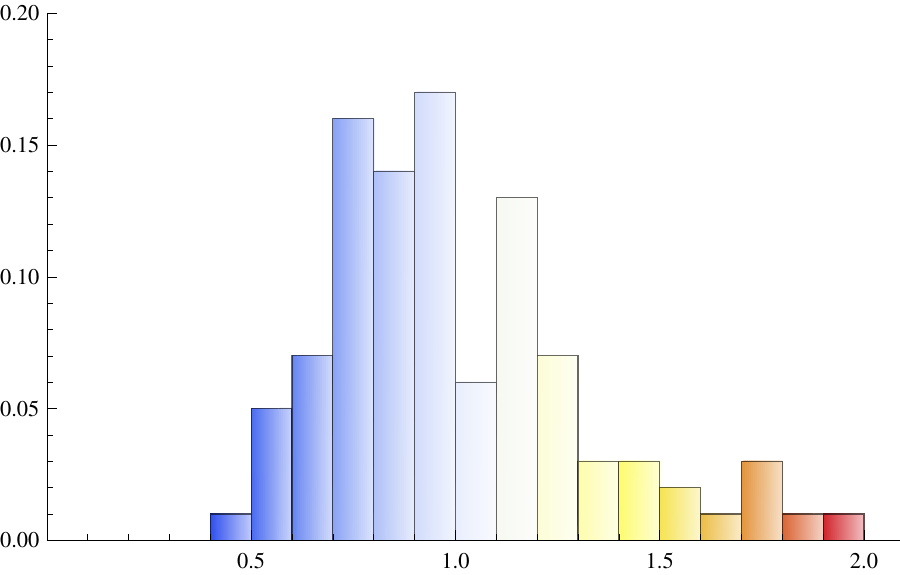}
	\end{subfigure}
	\\
	\begin{subfigure}{.328\textwidth}
		\raggedright
		\includegraphics[width=.95\linewidth]{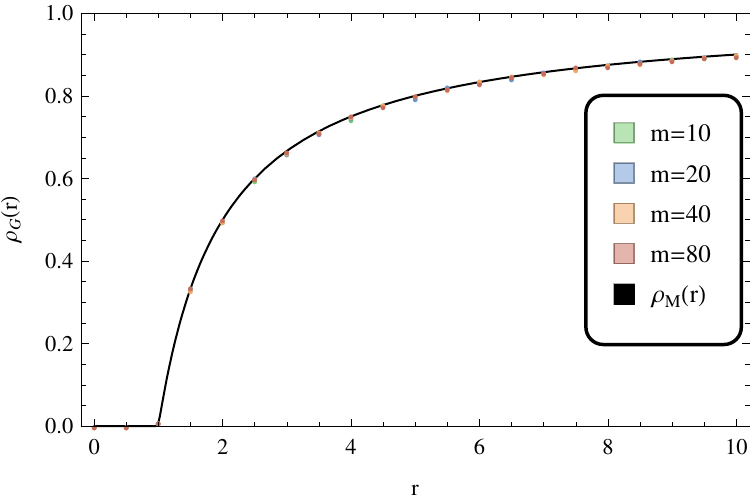}
	\end{subfigure}
	\begin{subfigure}{.328\textwidth}
		\raggedright
		\includegraphics[width=.95\linewidth]{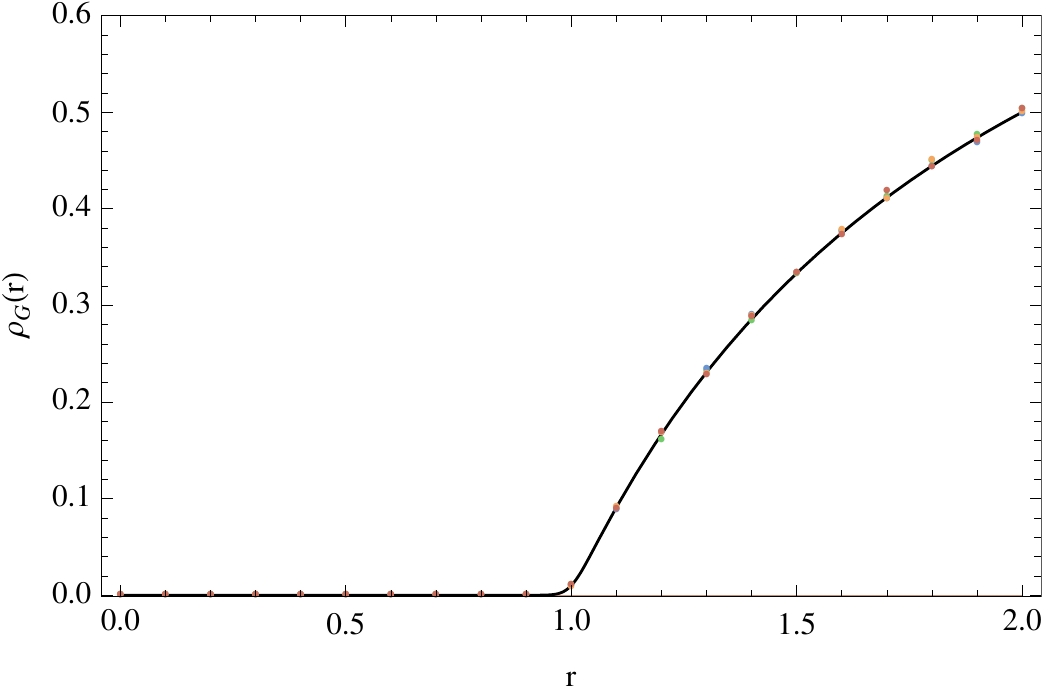}
	\end{subfigure}
	\begin{subfigure}{.328\textwidth}
		\raggedright
		\includegraphics[width=.95\linewidth]{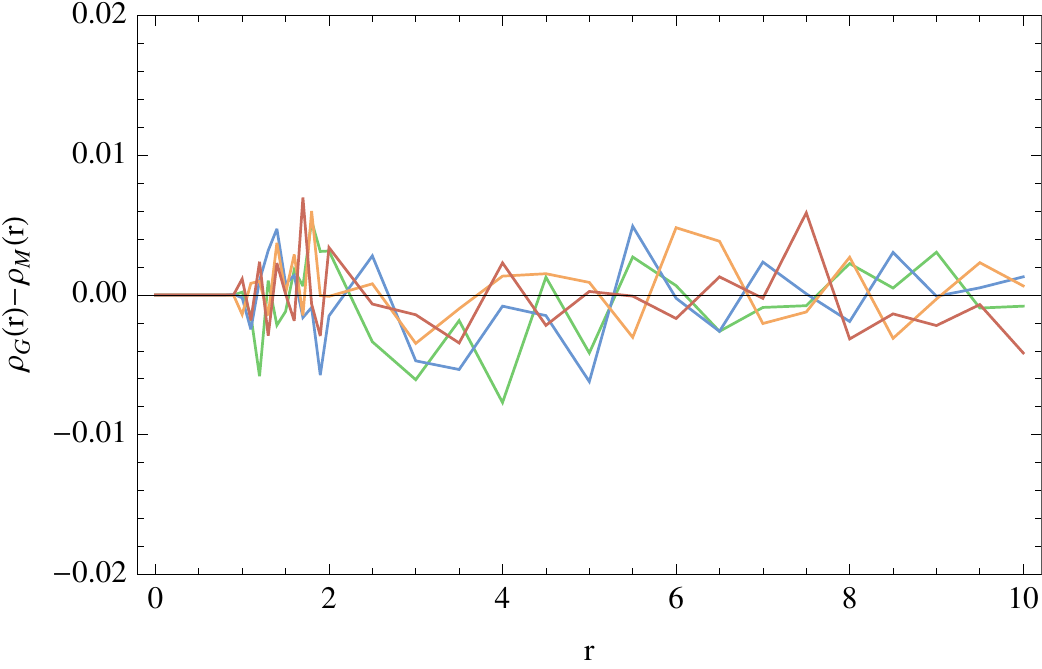}
	\end{subfigure}
	\caption{Simulations on graphs generated by preferential attachment yield fixation probabilities close to $\rhoM$. Several scale-free networks with varying out degrees, $m=10$, $m=20$, $m=40$, and $m=80$, were generated using a preferential attachment algorithm. Histograms of the sum of the weights of edges pointing to each vertex are plotted next to each graph, however, the small graphs size limits the resemblance to a power law. Given the comparatively large size of the graphs, only a restricted number of simulations were performed, but the simulations corresponded to $\rhoM$ without a tendency to amplify or repress. More extensive work is required.}
	\label{fig: sf}
\end{figure}

Surprisingly, even though there is a sense in which vertices are not treated interchangeably in the preferential attachment algorithm, Moran-type behavior was observed in all simulations (Figure \ref{fig: sf}). This is in contrast with the results in Lieberman \emph{et al.} where they observed some amplification in scale-free networks \cite{Lieberman2005}. The scale-free property is emergent and only becomes apparent as the graph becomes large, thus this increases the running time of the Monte Carlo method for estimating the fixation probability. More simulations are required here for conclusive findings and again there are currently no mathematical results.

In summary, we have generalized the isothermal theorem to make it biologically realistic and to increase its technical applicability. The conclusion of the robust isothermal theorem now depends continuously on its assumptions. With this new tool, we have proved analytically that fixation probabilities in a generalized Erd\H{o}s-R\'{e}nyi model converge uniformly to the fixation probability of a well-mixed population. In our proof, we identify the reason for this convergence and bound its rate. Thus, we confirm observations from many simulations and give a method of approximation with a specified error. Furthermore, we conjecture that many random graph models exhibit this universal behavior. However, it is easy to construct simple examples of random graphs which do not, thus it still remains to determine the necessary assumptions on the random graph model for it to exhibit universal behavior.

\newpage

\newpage

\newpage

\newpage

\newpage

\biblio


\appendix


\newpage
\maketitle

\section{The robust isothermal theorem}\label{sec: robust isothermal}

The Moran process on graphs is the standard model for population structure in evolutionary dynamics \cite{Lieberman2005, Nowak2006}. The process is defined for a directed, weighted graph, $G\equiv G_{n}=(V_{n},W_{n})$, where $V\equiv V_{n}\deq\db{n}$ and $W\equiv W_{n}\deq [w_{ij}]$ is a stochastic matrix of edge weights. The process $X_{t}$ is Markovian with state space $2^{V}$, where each $X_{t}$ is the subset of $V$ consisting of all the vertices occupied by mutants at time $t$. At time 0 a mutant is placed at one of the vertices uniformly at random or formally,
\eq{
	\P[X_{0}=S]=\begin{cases}
		n^{-1} & \text{if } |S|=1\\
		0 & \text{otherwise}
	\end{cases}.
}
Then at each subsequent time step exactly one vertex is chosen randomly, proportional to its fitness, for reproduction: so the probability of choosing a particular wild type vertex is $1/(n-|X_{t}|+r|X_{t}|)$ and the probability of choosing a particular mutant vertex is $r/(n-|X_{t}|+r|X_{t}|)$. An edge originating from the chosen vertex is then selected randomly with probability equal to its edge weight, which is well defined since $W$ is stochastic, and the vertex at the destination of the edge takes on the type of the vertex at the origin of the edge. 

Typically, there are exactly two absorbing states, $X_{t}=\emptyset$ and $X_{t}=V$, corresponding to the wild type fixing in the population and the mutant fixing in the population respectively. Thus, almost surely, one of these two absorbing states is reached in finite time. The probability that the process reaches $V$ and not $\emptyset$ is called the \emph{fixation probability} and for a graph $G$ we denote its fixation probability for a mutant of fitness $r>0$ by
\eq{
	\rho_{G}(r)\deq \P[ X_{t}=V \text{ for some } t\geq 0 ].
}

A fundamental point of comparison is the fixation probability $\rhoM$ for a \emph{well-mixed} population structure, where the graph structure is given by
\eq{
	w_{ij}\deq \begin{cases}
		(n-1)^{-1} & \text{if } i\neq j \\
		0 & \text{if } i=j
	\end{cases}
}
and M stands for ``Moran'' or ``mixed.'' An easy calculation using recurrence equations shows that
\eq{\label{eq: moran}
	\rhoM(r)=\frac{1-r^{-1}}{1-r^{-n}}.
}
Graphs with exactly the same fixation probability as $\rhoM$ are classified by the isothermal theorem, which gives sufficient conditions for a general graph $G$ to have the same fixation probability as $\rhoM$ \cite{Lieberman2005}.

In this section we derive a generalization of the isothermal theorem and throughout we shall require that the matrix $W$ is stochastic---that is, the row sums are all equal to 1:
\eq{
	\sum_{j=1}^{n}w_{ij}=1,
}
for all $i\in V$. Any graph with nonnegative edge weights can be normalized to produce a graph with a stochastic $W$, so long as each row has a nonzero entry, without changing the behavior of the process as defined above. A graph $G$ is called isothermal if all the column sums of $W$ are identical---that is, $W_{1}=\dotsb=W_{n}=1$, where
\eq{
	W_{j}\deq \sum_{i=1}^{n}w_{ij},
}
or equivalently, $W$ is doubly stochastic.  

For all $S\subseteq{V}$, define
\eq{
	w_{\text{O}}(S)\deq \sum_{i\in S}\sum_{j\not\in S}w_{ij} \quad\text{ and } \quad w_{\text{I}}(S)\deq \sum_{i\not\in S}\sum_{j\in S}w_{ij}
}
as the sum of the edge wights leaving $S$ and entering $S$ respectively. Then an easy calculation shows that a graph is isothermal if and only if 
\eq{\label{def: isothermal}
	w_{\text{O}}(S)=w_{\text{I}}(S)
}
for all $\emptyset\neq S\subsetneq V$. The later condition and its equivalence to isothermality is at the core of the proof of the isothermal theorem. The term ``isothermal'' originates from an interpretation of the sum of the ingoing edge weights as temperature, with ``hotter'' vertices changing more frequently in the Moran process. Thus a graph satisfying \eqref{def: isothermal} is isothermal because the ingoing and outgoing temperatures are equal and all subsets $S$ are in ``thermal equilibrium.'' We now restate the forward direction of the original isothermal theorem.

\begin{theorem}[Isothermal theorem]\label{thm: isothermal}
	Suppose that a graph $G$ is isothermal, then the fixation probability of a randomly placed mutant of fitness $r$ is equal to $\rhoM(r)$.
\end{theorem}

We ask, can we relax the assumptions of Theorem \ref{thm: isothermal}? That is, perhaps an approximate result can be obtained for $W$ that are only approximately doubly stochastic in the following sense:
\eq{
	\absa{W_{j}-1}\leq \e
}
for all $j\in V$ and some small quantity $\e$. 
However, the example
\eq{
	W=\left[\begin{array}{cccc}0 & 1-\e\ & \e & 0 \\1-\e & 0 & 0 & \e \\\e^2 & 0 & 0 & 1-\e^2 \\0 & \e^2 & 1-\e^2 & 0\end{array}\right]
}
shows we cannot, since as $\e\to0$ 
\eq{
	\frac{w_{\text{O}}(\ha{1,2})}{w_{\text{I}}(\ha{1,2})}=\frac{2\e}{2\e^{2}}=\e^{-1}\to\infty.
}
That is, $W$ is approximately doubly stochastic, but the ratio of the outgoing and ingoing edge weights is unbounded for some subset $S$. Thus, we need stronger assumptions for our theorem which we state now.

\begin{theorem}[Robust isothermal theorem]\label{thm: robust isothermal}
	Fix $0\leq \e<1$. Let $G_{n}=(V_{n},W_{n})$ be a connected graph. If for all nonempty $ S\subsetneq V_{n}$ we have
	\eq{\label{eq: robust assumption}
		\absa{\frac{w_{\text{O}}(S)}{w_{\text{I}}(S)}-1}\leq \e,
	}
	then
	\eq{\label{eq: robust isothermal}
		\sup_{r>0}\absa{\rhoM(r) - \rho_{G_{n}}(r)} \leq  \e.
	}
\end{theorem}

\begin{proof}
	To briefly outline the proof, we begin by projecting the process from $X_{t}$ to $\absa{X_{t}}$. Next we consider the ratio of the probability of increasing the number of mutants to the probability of decreasing the number of mutants. By bounding this ratio we can use a coupling argument to establish that the fixation probability of the process is close to $\rhoM$. Finally, we use the mean value theorem and smoothness properties of $\rhoM$ to simplify our bound and obtain the result.

	Just as in the proof of the original isothermal theorem, we make the projection of the state space of all subsets of $V$, which records exactly which vertices are mutants, to the simpler state space $\{0,1,\ldots,n\}$, which records only the number of mutants. The problem with making this projection in general is that the transition probabilities from one subset to another can depend on the structure of a subset not merely the number of mutants. However, it is clear that the only quantities which affect the fixation probability are the ratios of the probability of increasing the number of mutants to the probability of decreasing the number of mutants in a particular state $S$.
	
	Define $p_{+}(S)$ and $p_{-}(S)$ as the probability that the number of mutants in a population increases and decreases by one respectively. Thus
	\eq{
		p_{+}(S)=\frac{w_{\text{O}}(S)r}{w_{\text{O}}(S)r+w_{\text{I}}(S)} \quad \text{ and }\quad p_{-}(S)=\frac{w_{\text{I}}(S)}{w_{\text{O}}(S)r+w_{\text{I}}(S)},
	}
	which gives, when the two equations are divided,
	\eq{\
		\frac{ p_{+}(S) }{ p_{-}(S) }=r\frac{ w_{\text{O}}(S) }{ w_{\text{I}}(S) }.
	}
	By assumption \eqref{eq: robust assumption},
	\eq{\label{eq: ratio bound}
		r(1-\e) \leq \frac{ p_{+}(S) }{ p_{-}(S) } \leq r(1+\e).
	}
	This states that the ratio of the probabilities of increasing to decreasing the number of mutants in any state $S$ is approximately proportional to $r$. 
	
	If for some graph $G'=(V',W')$ we have $ p_{+}(S) / p_{-}(S)  = r(1\pm\e)$ for all $S\subseteq V'$, then by the standard result for fixation probabilities in birth-death processes, its fixation probability is given by
	\eq{\label{eq: bdc result}
		\rho_{G'}(r)=\rhoM(r\pm r\e)=\frac{1-(r\pm\e r)^{-1}}{1-(r\pm\e r)^{-n}}.
	}
	From \eqref{eq: ratio bound} and \eqref{eq: bdc result} we would like to conclude that 
	\eq{\label{eq: bound 1}
		\rhoM(r-r\e)=\frac{1-(r-\e r)^{-1}}{1-(r-\e r)^{-n}} \leq \rho_{G}(r) \leq \frac{1-(r+\e r)^{-1}}{1-(r+\e r)^{-n}}=\rhoM(r+r\e ).
	}
	The upper bound is given by taking the maximum allowed value for the probability of increasing the number of mutants relative to the probability of decreasing the number of mutants. For the lower bound we use the opposite. 
	
	This intuitive result can be proved with a coupling argument. We can couple the Moran process $X_t$ of a mutant of fitness $r$ on $G$ with another process $Y$ defined as follows: $Y$ has state space $\{0,\ldots, n\}$ (with 0 and $n$ absorbing) and $Y$ starts at $1$. We couple $Y$ to $X_t$ as follows: 
	\begin{enumerate}
		\item if $\absa{X_t}$ decreases by 1, then $Y$ must also decrease by 1;
		\item if $\absa{X_t}$ increases by 1, then independently $Y$ increases by 1 with probability
	\eq{
		\frac{P_+(S)+P_-(S)}{P_+(S)}\frac{r(1-\e)}{1+r(1-\e)}
	}
	(which is less than or equal to 1 by assumption \eqref{eq: ratio bound}), else $Y$ decreases by 1; 
		\item otherwise $Y$ remains constant.
	\end{enumerate}
	Note that marginally $Y$ is a simple random walk on $\{0,\ldots, n\}$ with forward bias $r(1-\e)$ and thus the probability that $Y$ reaches $n$ before it reaches $0$ is given by
	\eq{\label{eq: Y fix}
		\frac{1-(r-\e r)^{-1}}{1-(r-\e r)^{-n}}.
	}
	However, because the processes are coupled we have $Y\leq |X_t|$ and thus if $Y=n$, then the mutant has fixed in the process $X_t$. 	Equation \eqref{eq: Y fix} immediately implies the lower bound in \eqref{eq: bound 1}. A similar coupling yields the upper bound. Thus 
	\eq{
		\rhoM(r-r\e) -\rhoM(r) \leq \rho_{G}(r) -\rhoM(r)\leq \rhoM(r+r\e)-\rhoM(r).
	}

	By the mean value theorem, 
	\eq{
		\frac{\rhoM(r+r\e )-\rhoM(r)}{\e } \leq \frac{r \e}{\e}\sup_{r\leq x\leq r+r\e}\absa{\rhoM'(x)}=r\sup_{r\leq x\leq r+r\e}\absa{\rhoM'(x)}
	}
	and
	\eq{
		\frac{\rhoM(r)-\rhoM(r-r\e)}{\e} \geq -r\sup_{r-r\e_{n}\leq x\leq r}\absa{\rhoM'(x)}.
	}
	Thus, it is sufficient to show for all $r>0$
	\eq{
		\sup_{n\geq 2}\absa{\rhoM'(r)}\leq r^{-1}.
	}
	We note that this is not an optimal bound, however, it sufficies for our applications. Calculating, one finds
	\eq{
		\rhoM'(r)=\frac{r^{n-2} \left(r^n-n r+n-1\right)}{\left(r^n-1\right)^2}.
	}
	First, when $r\geq1$ we prove the stronger claim 
	\eq{\label{eq: derivative bound 1} 
		\frac{r^{n-2} \left(r^n-n r+n-1\right)}{\left(r^n-1\right)^2}\leq r^{-2},
	}
	by noting the above is equivalent to
	\eq{
		(r-1)\pa{nr^{n}-\sum_{k=0}^{n-1}r^{k}}\geq 0,
	}
	which is true since $r\geq1$. Similarly, one can prove
	\eq{\label{eq: derivative bound 2}
		\frac{r^{n-2} \left(r^n-n r+n-1\right)}{\left(r^n-1\right)^2}<1
	}
	when $r<1$. Equations \eqref{eq: derivative bound 1} and \eqref{eq: derivative bound 2} imply $\sup_{n\geq 2}\absa{\rhoM'(r)}\leq r^{-1}$.
	
	Therefore, we may conclude
	\eq{
		\sup_{r>0}\absa{\rho_{G}(r) -\rhoM(r)}\leq \e,
	}
	which completes the proof.
\end{proof}

	Note that one can sometimes do slightly better (depending on the relative sizes of $\e$ and $n^{-1}$), by showing 
	\eq{
		\sup_{r\leq 1}\absa{\rho_{G_{n}}(r) -\rhoM(r)}< \frac{2}{n}
	}
	for $\e$ small enough, using the triangle inequality and the facts that $\rhoM$ is increasing and continuous, but again this is not important for our applications.

\begin{figure}[ht!]
	\centering
	\includegraphics[width=.5\linewidth]{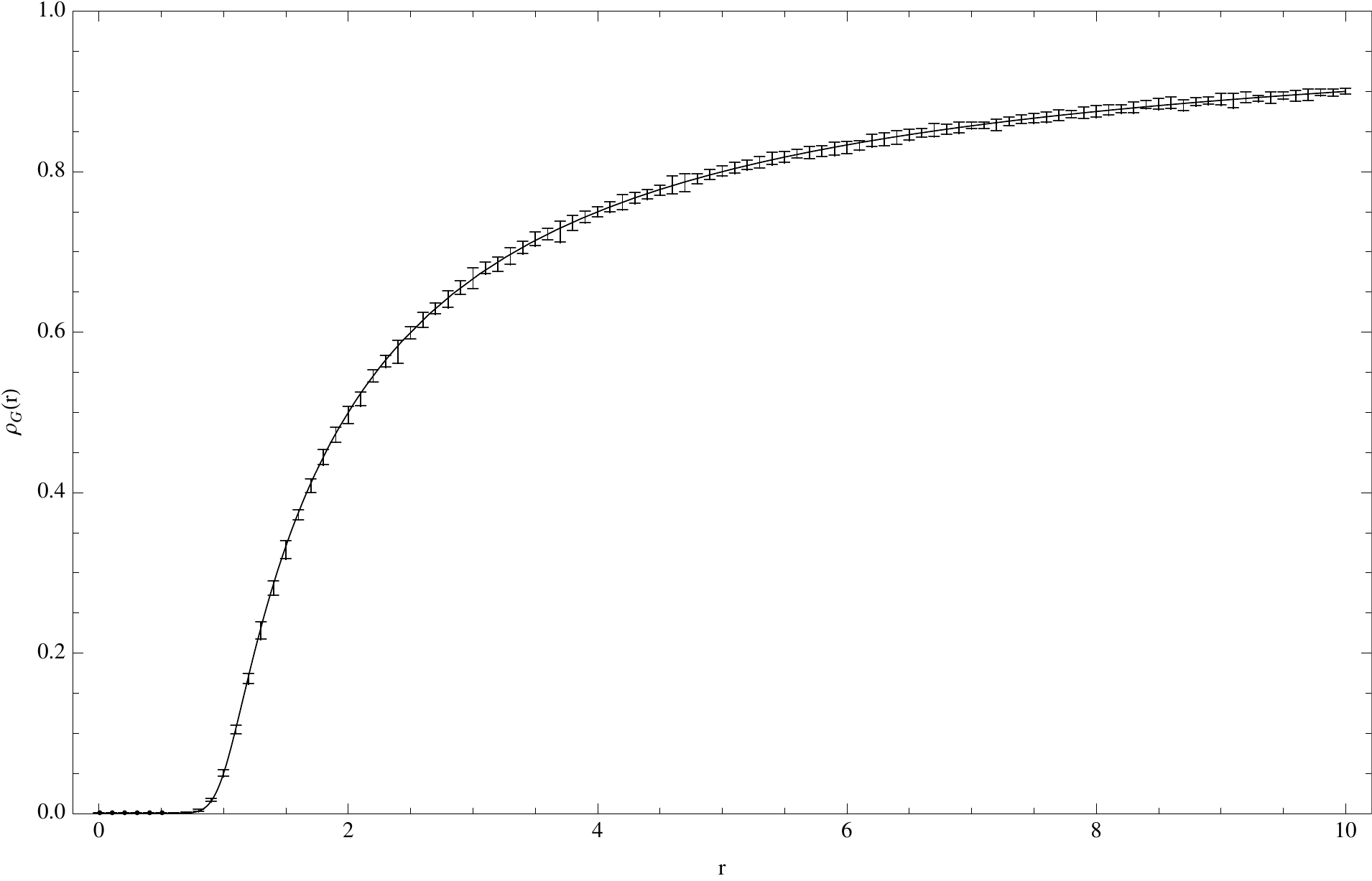}
	\caption{Fixation probability does not depend on starting location. We conducted trials where the fixation probability of a mutant of fitness $r$ starting at vertex $i$ was estimated with the Monte Carlo method of $10^{4}$ samples for several values of $0\leq r\leq 10$ on a Erd\H{o}s-R\'{e}nyi random graph. The fixation probability was similar regardless of starting vertex, and in particular, showed no correlation with vertex temperature. We plot the Moran fixation probability \eqref{eq: moran} and use the error bars to illustrate the minimum and maximum empirical fixation probabilities obtained from starting at any particular vertex.}
	\label{fig: start}
\end{figure}

Theorem \ref{thm: robust isothermal} is actually slightly stronger than stated, and thus we can draw a slightly stronger conclusion in Theorem \ref{thm: random graph}. We may conclude that the fixation probability of a mutant of fitness $r$ originating at a particular vertex $i$ \cite{Antal2006} satisfies the bound in \eqref{eq: robust isothermal}, for exactly the same reason as in the proof of the original isothermal theorem---the bound \eqref{eq: robust assumption} is for all subsets $S$. Therefore, \emph{a fortiori}, a mutant can be started with any probability vector on the vertices (not merely uniform) and its fixation probability will still satisfy \eqref{eq: robust isothermal}. This observation is borne through simulations too (Figure \ref{fig: start}). 

\section{Evolution on random graphs}\label{sec: random graphs}

In this section we prove Theorem \ref{thm: random graph}. 

\begin{theorem}\label{thm: random graph}
	Let $\pa{G_{n}}_{n\geq 1}$ be a family of random graphs as in Definition \ref{def: random graph}. Then there are constants $C>0$ and $c>0$, not dependent on $n$, such that the fixation probability of a randomly placed mutant of fitness $r>0$ satisfies
	\eq{\label{eq: random graphs 1}
		\absa{\rho_{G_{n}}(r)-\rhoM(r)}\leq  \frac{C\pa{\log n}^{C+C\xi}}{\sqrt{n}}
	}
	uniformly in $r$ with probability greater than
	\eq{
		1- \exp\pa{-\nu \pa{\log n}^{1+\xi}},
	}
	for positive constants $\xi$ and $\nu$.
\end{theorem}

To do this we need to apply Theorem \ref{thm: robust isothermal} to our random graphs by showing that its assumptions hold with high probability. We do so in several steps. First, we define precisely generalized Erd\H{o}s-R\'{e}nyi random graphs in Definition \ref{def: random graph} and outline the necessary assumptions on the distribution of the edge weights. After reviewing some notation, we introduce an event $\Omega$, on which the graphs are well behaved, and show that $\Omega$ has high probability in Lemma \ref{lem: Omega is hp}.  Then the general idea is to use large deviation estimates and concentration inequalities to show that with high probability the quantity \eqref{eq: robust assumption} can be controlled. We bound both the numerator (Lemma \ref{lem: column bounds}) and denominator (Lemma \ref{lem: weight bounds}) of
\eq{
	 \frac{\absa{w_{\text{O}}(S)-w_{\text{I}}(S)}}{\absa{w_{\text{I}}(S)}}=\absa{\frac{w_{\text{O}}(S)}{w_{\text{I}}(S)}-1}
}
for all $S$, then we put everything together to prove Theorem \ref{thm: random graph}.

\begin{remark}[Notation]\label{rem: constants and order notation}
	We use the large constant $C>0$ and the small constant $c>0$, which do not depend on the size of the graph $n$ but can depend on the distribution as outlined in Definition \ref{def: random graph}. We allow the constant $C$ to increase or the constant $c$ to decrease from line to line without noting it or introducing a new notation, sometimes we even absorb other constants such as $p$, $p'$, and $\mu_{1}$ without noting it; as is clear from the proof, this only happens a finite number of times, and thus we end with constants $C>0$, $c>0$, $\nu>0$, and $\xi>0$.
	
	We also make use of standard order notation for functions, $o(\cdot)$, $\cal{O}(\cdot)$, and $\cdot\gg\cdot$, all of which are used with respect to $n$. Moreover, in some sums it is useful to exclude particular summands, e.g.
	\eq{
		\sum_{\substack{1\leq j \leq n \\ j\not\in S}}\cdot \equiv \sum_{j}^{(S)}\cdot
	}
	for $S\subset V$. We abbreviate $(\{i\})$ and $(\{i,k\})$ as $(i)$ and $(i,k)$.
\end{remark}

\begin{remark}[High probability events]\label{rem: high probability}
	We say that an $n$-dependent event $E$ holds with high probability if, for constants $\xi>0$ and $\nu>0$ which do not depend on $n$, 
	\eq{
		\P[E^{c}]\leq e^{-\nu \pa{\log n}^{1+\xi} }
	}
	for $n\geq n_{0}\pa{ \nu, \xi }$. Moreover, we say an event $E$ has high probability on another event $E_{0}$ if
	\eq{
		\P[E_{0}\cap E^{c}]\leq e^{-\nu \pa{\log n}^{1+\xi} }
	}
	In particular, this has the property that the intersection of polynomially many (in $n$, say $Kn^{K}$ for some constant $K>0$) events of high probability is also an event of high probability: by the union bound,
\eq{
	\P\qa{\pa{\bigcap_{i=1}^{Kn^{K}}E_{i}}^{c}}=\P\qa{\bigcup_{i=1}^{Kn^{K}}E_{i}^{c}}\leq Kn^{K}\cdot e^{-\nu \pa{\log n}^{1+\xi} }=Ke^{K\log n-\nu\pa{\log n}^{1+\xi}}\leq e^{-\nu\pa{\log n}^{1+\xi}},
}
with a possible increase in $ n_{0}\pa{ \nu, \xi }$ and a decrease in the constants $\nu$ and $\xi$.
\end{remark}

\subsection{Proof of Theorem \ref{thm: random graph}} Following the Erd\H{o}s-R\'{e}nyi model, we produce a weighted, directed graph as follows: Consider an $n\times n$ matrix $X=[x_{ij}]$ with zero for its diagonal entries and independent, identically distributed, nonnegative random variables for its off-diagonal entries. We now want to define a random, stochastic matrix $W$ of weights. The natural definition for $W=[w_{ij}]$ is
\eq{\label{eq: basic def}
	w_{ij}\deq \frac{x_{ij}}{\sum_{k=1}^{n}x_{ij}},
}
which is defined when at least one of the $x_{i1},\ldots, x_{in}$ is nonzero; this happens almost surely in the limit as $n\to\infty$, when $\P[x_{ij}>0]=p>0$ is a constant:
\eq{
	\P[ x_{i1}=x_{i2}=\dotsb=x_{in}=0]=(1-p)^{n-1}
}
and by the union bound
\eq{
	\P\qa{ \bigvee_{i=1}^{n}x_{i1}=x_{i2}=\dotsb=x_{in}=0}\leq \sum_{i=1}^{n}\P[ x_{i1}=x_{i2}=\dotsb=x_{in}=0]=n(1-p)^{n-1}\to0.
}

However, the question is how to technically deal with the unusual event that all the entries of a row of $X$ are zero, as there are several options. We make the following choice: for $1\leq i\leq n$
\eq{\label{eq: w def 1}
	w_{ii}\deq \begin{cases}
		1 & \text{if } x_{i1}=x_{i2}=\dotsb=x_{in}=0\\
		0 & \text{otherwise}
	\end{cases},
}
and for all $1\leq i,j\leq n$ and $i\neq j$
\eq{\label{eq: w def 2}
	w_{ij}\deq \begin{cases}
		\frac{x_{ij}}{\sum_{k=1}^{n}x_{ik}} & \text{if } x_{ij}>0\\
		0 & \text{if } x_{ij}=0
	\end{cases}.
}
This definition aligns with the definition in \eqref{eq: basic def} with probability greater than $1-n(1-p)^{n}$. Moreover, this definition has the advantage that the events that any non-loop edge weight is 0 are independent.

\begin{definition}[Generalized Erd\H{o}s-R\'{e}nyi random graphs]\label{def: random graph}
	Let $\mu$ be a nonnegative distribution (not depending on $n$) with subexponential decay such that if $X\sim\mu$
	\eq{\label{eq: distribution assumptions}
		\P[X>0]= p>0 ~ \text{and} ~ \P\qa{ X\geq x }\leq Ce^{-x^{c}}
	}
	for some positive constants $p$, $c$, and $C$ and all $x>0$. We denote the mean and standard deviation of $X$ by $\mu_{1}$ and $\sigma$ respectively. We generate a family of random graphs $G_{n}=(V_{n},W_{n})$ from $\mu$ by defining the weight matrices $W_{n}$ according to \eqref{eq: w def 1} and \eqref{eq: w def 2}, where $x_{ij}$ are independent and distributed according to $\mu$ for $i\neq j$.
\end{definition}

The subexponential decay is necessary to control the fluctuation of the graph's edge weights and imposes a bounded increase on the moments of $\mu$. Let $X\sim \mu$, then simple calculations show,
\eq{\label{eq: moments of mu}
	\mu_{k}\deq \E X^{k} \leq C k \int_{0}^{\infty}x^{k-1} \P[X\geq x]\dd x \leq Ck \int_{0}^{\infty}x^{k-1} e^{-x^{c}}\dd x =\frac{C}{c}k\Gamma\pa{\frac{1+k}{c}}\leq \pa{Ck}^{Ck},
}
where the constant $C>0$ depends on the constants in \eqref{eq: distribution assumptions}. Many distributions satisfy the subexponential assumption \eqref{eq: distribution assumptions}, for example any compactly supported distribution, the Gamma distribution, and the absolute value of a Gaussian distribution. 

We now use the subexponential decay assumption to understand the typical behavior of the random variables $x_{ij}$. 

\begin{definition}[Good events $\Omega$]\label{def: Omega}
	Let $\Omega$ be an $n$-dependent event such that the following hold:
	\al{
		\Omega\deq \bigcap_{i=1}^{n}&\pa{\ha{x_{ii}=0 }  \cap \ha{ \absa{ \sum_{j}^{(i)}\pa{x_{ij}-\E x_{ij}} }\leq \sigma\pa{\log n}^{C+C\xi} \sqrt{n}  }} \\
		&\cap \bigcap_{i,j=1}^{n} \ha{ x_{ij}\leq C\pa{\log n}^{C}} \cap\ha{ G_{n} \text{ is connected} }.
	}
\end{definition}

The conditions on $\Omega$ have natural interpretations. The first condition specifies that the normalization procedure outlined above has worked as intended and that we are not in the atypical case where the graph has a self-loop. The second condition specifies that the sums of $n$ of the $x_{ij}$s are close to their expectation $(n-1)\mu_{1}$ and that they fluctuate about this value on the order of $\sqrt{n}$ as predicted by the central limit theorem. The third condition says that none of the $x_{ij}$ are too large and that typically they will all be less than $C\pa{\log n}^{C}$. The last condition is self-explanatory, as the Moran process is not guaranteed to terminate on disconnected graphs.

\begin{lemma}\label{lem: Omega is hp}
	The event $\Omega$ holds with high probability.
\end{lemma}

\begin{proof}
By Remark \ref{rem: high probability}, it suffices to show that each conjunct holds with high probability as there are only polynomially many choices for $i$ and $j$. First fix $i$. By assumption \eqref{eq: distribution assumptions} and the fact that $x_{ii}\neq0$ only if $x_{ij}=0$ for all $j\neq i$, 
\eq{
	\P\ha{x_{ii}\neq0} = \P\qa{ x_{ij}=0 \text{ for all } j\neq i }\leq (1-p)^{n-1} =e^{\log (1-p)(n-1)}\leq e^{-\nu \pa{\log n}^{1+\xi} },
}
since $0<p<1$ and $n-1\gg (\log n)^{1+\xi}$. 

Now using the large deviation result, Lemma \ref{lem: lde}, with $a_{j}=x_{ij}-\E x_{ij}$ and $A_{j}=1$, we may verify the moment assumption \eqref{eq: lde moment conditions}: clearly $\E \pa{x_{ij}-\E x_{ij}}=0$ and $\E \pa{x_{ij}-\E x_{ij}}^{2}=\sigma^{2}$, then
\eq{
	\E \absa{x_{ij}-\E x_{ij}}^{k}\leq \pa{Ck}^{Ck}
}
by Equation \eqref{eq: moments of mu}. Thus we get
\eq{
	\P\qa{ \absa{ \sum_{i=1}^{n} a_{i}A_{i}  } \geq \sigma\pa{\log n}^{C+C\xi} \sqrt{n} } \leq e^{-\nu\pa{\log n}^{1+\xi}}.
} 

Now fix $j$ too. Next, we use the subexponential decay assumption \eqref{eq: distribution assumptions} with $x=C\pa{\log n}^{C}$ for $C=1+c^{-1}$ to get
\eq{
	\P\qa{x_{ij}> C\pa{\log n}^{c^{-1}+1}}\leq C\exp\pa{ -\pa{C\pa{\log n}^{c^{-1}+1}}^{c} }\leq Ce^{-C^{c}\pa{\log n}^{1+c}}\leq e^{-\nu \pa{\log n}^{1+\xi} }.
}	
Thus, $x_{ij}> C\pa{\log n}^{C}$ holds with high probability since $c>0$ and $C^{c}>0$.

Finally, we show that the graph $G$ is connected with high probability, i.e. that with high probability, the graph cannot be partitioned into two disjoint sets where there are no edges going from one subset to another. This follows from an argument similar to that contained in the proof of Lemma \ref{lem: weight bounds} but without the assumption that we are on the event $\Omega$ as we do not need a lower bound on the weights only that edges exist which they do with probability at least $p$.
\end{proof}

Note that by definition $W$ is stochastic. Define the sum of the $j$th column as
\eq{
	W_{j}\deq\sum_{i=1}^{n}w_{ij}.
}
Note that while the family $W_{j}$ is not independent, by symmetry, they are identically distributed. Hence
\eq{
	\E W_{1}=\frac{1}{n}\sum_{j=1}^{n}\E W_{j}=\frac{1}{n}\E \sum_{j=1}^{n}\sum_{i=1}^{n}w_{ij}=1.
}
This tells us that in expectation $W$ is doubly stochastic. The next lemma shows that with very high probability it is almost $n^{-1/2}$ close to being doubly stochastic, which is exactly the order of fluctuation we expect by the central limit theorem. The assumptions on the distribution $\mu$ and the event $\Omega$ guarantee that we can prove that the sum's fluctuations are of this order.

The idea of the proof is that for fixed $j$, the $w_{ij}$ are independent random variables and thus we can apply a {\sc lde} to bound the fluctuations of their sum. There are complications due to the normalization required by Definition \ref{def: random graph} but on $\Omega$ these can be overcome by relating the sum $W_{j}$ to a simpler sum that may be controlled with Lemma \ref{lem: lde}. 

\begin{lemma}\label{lem: column bounds}
	On $\Omega$, there are positive constants $c\equiv c_{\mu}$ and $C\equiv C_{\mu}$, not dependent on $n$, such that the following inequalities hold
	\eq{
		\absa{ W_{j} -1}\leq \frac{C\pa{\log n}^{C+C\xi}}{\sqrt{n}}
	}
	for all $j\in V$, with probability at least
	\eq{
		1-e^{-\nu \pa{\log n}^{1+\xi}}.
	}
\end{lemma}
\begin{proof}
Fix $j$. First we use the fact that $w_{ii}=x_{ii}=0$ for $1\leq i\leq n$ on $\Omega$ to see
\eq{
	 W_{j}-1=\sum_{i} \pa{w_{ij}-\E w_{ij}} = \sum_{i}^{(j)}\pa{w_{ij}-\frac{1}{n-1}}+ \cal{O}\pa{n^{2}(1-p)^{-n+1}}.
}
By the definition of $w_{ij}$, the above is equal to
\eq{\label{eq: step 2}
	\sum_{i}^{(j)}\pa{\frac{x_{ij}}{\sum_{k}^{(i)}x_{ik}}-\frac{1}{n-1}}+ \cal{O}\pa{c_{0}^{-n}}=\sum_{i}^{(j)}\pa{  \frac{x_{ij} -\frac{1}{n-1}\sum_{k}^{(i)}x_{ik} }{\sum_{k}^{(i)}x_{ik} }  }+ \cal{O}\pa{c_{0}^{-n}},
}
where $c_{0}<1$ is not dependent on $n$. Next, using the fact that on $\Omega$
\eq{\label{eq: average bound}
	\frac{1}{n-1}\absa{ \sum_{k}^{(i)}\pa{x_{ik}-\E x_{ik}} }\leq C\sigma\pa{\log n}^{C+C\xi} \frac{1}{\sqrt{n}}
}
for all $1\leq i\leq n$, we replace the average in the numerator of \eqref{eq: step 2} with its expectation to find it equal to
\eq{\label{eq: step 3}
	\sum_{i}^{(j)}\pa{  \frac{x_{ij} -\E x_{ij} }{\sum_{k}^{(i)}x_{ik} } + \frac{C\sigma\pa{\log n}^{C+C\xi}}{\sqrt{n}\sum_{k}^{(i)}x_{ik}  } }+ \cal{O}\pa{c_{0}^{-n}}.
}
Using \eqref{eq: average bound} again, it is easy to see
\eq{\label{eq: lower bound on denominator sum}
	 \sum_{k}^{(i)}x_{ik} \geq (n-1)\E x_{ij}-C\sigma\pa{\log n}^{C+C\xi}\sqrt{n},
}
which gives an upper bound on the error term in \eqref{eq: step 3} and we find the equation equal to
\eq{\label{eq: step 4}
	\sum_{i}^{(j)}  \frac{x_{ij} -\E x_{ij} }{\sum_{k}^{(i)}x_{ik} } + \cal{O}\pa{ \frac{\pa{ \log n }^{2C+2C\xi} }{\sqrt{n}} }  + \cal{O}\pa{c_{0}^{-n}}.
}
Next we compare these two expressions to find that the absolute value their difference can be expressed as
\eq{
	\absa{\frac{x_{ij} -\E x_{ij} }{\sum_{k}^{(i)}x_{ik} } - \frac{x_{ij} -\E x_{ij} }{\sum_{k}^{(i,j)}x_{ik} } } = \frac{ \absa{x_{ij}}^{2} }{ \absa{\sum_{k}^{(i)}x_{ik}\cdot \pa{\sum_{k}^{(i)}x_{ik} -x_{ij}} }}.
}
However, using that on $\Omega$, for all $1\leq i,j \leq n$, we have $x_{ij}\leq C\pa{\log n}^{C}$ and using \eqref{eq: average bound} as before, we may show the difference is bounded by
\eq{
	\cal{O}\pa{\frac{\pa{\log n}^{4C+2C\xi}}{n^{2}}}.
}
We can then sum over these errors---one for each summand---to get a total error of $\cal{O}\pa{\pa{\log n}^{4C+2C\xi}/n}$. Thus, \eqref{eq: step 4} may be rewritten as 
\eq{\label{eq: step 5}
	\sum_{i}^{(j)}  \frac{x_{ij} -\E x_{ij} }{\sum_{k}^{(i,j)}x_{ik} } + \cal{O}\pa{ \frac{\pa{ \log n }^{2C+2C\xi} }{\sqrt{n}} } ,
}
since the other error terms are dominated by the remaining one. 

Note that $x_{ij}$ does not appear in the summand's denominator and thus the denominator and numerator are independent. So we can use the large deviation estimate, Lemma \ref{lem: lde}, with $a_{i}=x_{ij}-\E x_{ij}$ and $A_{i}^{-1}=\sum_{k}^{(i,j)}x_{ik}$. While the $A_{i}$ are random, we may condition on their values and treat them as deterministic constants, then after we have used the {\sc lde}, we can bound them using the fact that we are on $\Omega$. That is, on $\Omega$
\eq{
	\pa{\sum_{k}^{(i,j)}x_{ik}}^{2}=(n-2)^{2}\pa{\E x_{ij}}^{2}+\cal{O}\pa{ \pa{\log n}^{2C+2C\xi}n\sqrt{n} }
}
and so
\eq{
	\sum_{k}^{(j)}A_{i}^{2}=\frac{1}{(n-2)\pa{\E x_{ij}}^{2}} + \cal{O}\pa{ \pa{\log n}^{2C+2C\xi}\frac{1}{n\sqrt{n}} }.
}
Thus the {\sc lde} gives us
\eq{
	\P\qa{ \absa{ \sum_{i}^{(j)} a_{i}A_{i}  } \geq \frac{C\pa{\log n}^{C+C\xi}}{\sqrt{n}}  } \leq e^{-\nu\pa{\log n}^{1+\xi}},
}
which combined with \eqref{eq: step 5}
\eq{
	\P\qa{ \absa{ W_{j} -1 } \geq \frac{C\pa{\log n}^{2C+2C\xi}}{\sqrt{n}}  } \leq e^{-\nu\pa{\log n}^{1+\xi}}.
}
The properties of high probability and the fact that we have $n$ choices for $j$ completes the proof.
\end{proof}

Next we prove a lower bound on sums of edge weights, $w_{\text{I}}(S)$ and $w_{\text{O}}(S)$ for all $\emptyset\neq S\subsetneq V_{n}$. The proof relies on concentration inequalities for independent random variables and the simple fact that on $\Omega$ there is a constant $c>0$ such that $w_{ij}\geq cn^{-1} \I\pa{ x_{ij}\geq c }$ for all $i,j\in V$.

\begin{lemma}\label{lem: weight bounds}
	On $\Omega$, for all $\emptyset\neq S\subsetneq V_{n}$ and some small constant $c\equiv c_{\mu}>0$, not dependent on $n$, we have the following bound
	\eq{
		|w_{\text{I}}(S)|=|w_{\text{O}}(S^{c})| \geq c_{\mu} \min\ha{|S|,n-|S|}
	}
	with probability greater than
	\eq{
		1-e^{-\nu\pa{\log n}^{1+\xi}}.
	}
\end{lemma}
\begin{proof}
First note that as in the proof of Lemma \ref{lem: column bounds}, we can argue that on $\Omega$ the sum $\sum_{k}^{(i)}x_{ik}\leq Cn $, see \eqref{eq: lower bound on denominator sum} for all $1\leq i\leq n$. Moreover, by assumption on the distribution $\mu$, we have $\P\qa{ x_{ij}>0 }=p>0$ and thus there is a constant $c>0$ such that $\P\qa{ x_{ij}\geq c }=p'>0$. Therefore, on $\Omega$
\eq{
	w_{ij}\geq cn^{-1} \I\pa{ x_{ij}\geq c }.
}
However, for each $1\leq i,j \leq n$ with $i\neq j$, define $\beta_{ij}\deq\I\pa{ x_{ij}\geq c }$ which are independent Bernoulli random variables such that
\eq{\label{eq: bernoulli lower bound}
	\P[\I\pa{ x_{ij}\geq c }=1]=p'>0,
}
since the $x_{ij}$ are independent.

Let $|S|=k$. By definition
\eq{
	w_{\text{I}}(S)=\sum_{i\not\in S}\sum_{j\in S}w_{ij} =w_{\text{O}}(S^{c}).
}
Note that no diagonal terms are in these sums. Using \eqref{eq: bernoulli lower bound},
\eq{
	\sum_{i\in S}\sum_{i\not\in S}w_{ij} \geq \frac{c}{n}\sum_{i\in S}\sum_{i\not\in S}\beta_{ij}.
}
Note that now this is a sum of $k(n-k)$ independent random variables. So for fixed $\emptyset\neq S\subsetneq V_{n}$, by the Chernoff bound, Lemma \ref{lem: mcb}, the event
\eq{
	A_{S}\deq \ha{ \sum_{i\in S}\sum_{i\not\in S}\beta_{ij}\leq (1-1/2)p' k(n-k) }
}
has probability less than
\eq{
	\exp\pa{-\frac{1}{8}p'k(n-k)}.
}
Thus, by the union bound
\al{\begin{split}
	\P\qa{ \bigcap_{\emptyset\neq S\subsetneq V_{n}} A_{S}^{c}}&=1-\P\qa{ \bigcup_{\emptyset\neq S\subsetneq V_{n}} A_{S}}\\
	&\geq 1- \sum_{\emptyset\neq S\subsetneq V_{n}}\P\qa{A_{S}}\\
	&= 1-\sum_{k=1}^{n-1}\binom{n}{k}\P\qa{A_{S}} \\
	&=1-\sum_{k=1}^{\floor{\log n}}\binom{n}{k}\P\qa{A_{S}}-\sum_{k=\floor{\log n}+1}^{n-\floor{\log n}-1}\binom{n}{k}\P\qa{A_{S}}-\sum_{n-\floor{\log n}}^{n-1}\binom{n}{k}\P\qa{A_{S}}\\
	&\geq1-2n^{\log n}\exp\pa{ -p'n/8 }-2^{n}\exp\pa{-p'n\pa{\log n}/8}\\
	&\geq 1-2\exp\pa{ -\pa{p'n/8-(\log n)^{2} }}-\exp\pa{ -\pa{p'\log n/8-\log 2}n }\\
	&\geq 1- \exp\pa{ -cp'n },
\end{split}}
for some $c>0$. Finally, note 
\eq{
	\frac{cp'}{2n}k(n-k)\geq\frac{cp'}{2} \min\ha{|S|,n-|S|}
}
and 
\eq{
	\exp(-cp'n) \leq e^{-\nu\pa{\log n}^{1+\xi}}
}
for an appropriate choice of $\xi$ and $\nu$.
\end{proof}

We now complete the proof of Theorem \ref{thm: random graph} by putting together the results of Section \ref{sec: robust isothermal} and the lemmata from this section.

\begin{proof}[Proof of Theorem \ref{thm: random graph}]
Again let $|S|=k$. We check that the assumptions of Theorem \ref{thm: robust isothermal} hold with high probability. Observe
\eq{
	\absa{\frac{w_{\text{O}}(S)}{w_{\text{I}}(S)}-1}=\absa{ \frac{w_{\text{O}}(S)-w_{\text{I}}(S)}{w_{\text{I}}(S)} }=  \frac{\absa{w_{\text{O}}(S)-w_{\text{I}}(S)}}{\absa{w_{\text{I}}(S)}}.
}
Expanding the numerator, we get
\eq{\label{eq: sum expanded 1}
	w_{\text{O}}(S)-w_{\text{I}}(S)=\sum_{i\in S}\sum_{j\not\in S} w_{ij} -\sum_{i\not\in S}\sum_{j\in S}w_{ij}=\sum_{i\in S}\sum_{j\in V}w_{ij}-\sum_{i\in V}\sum_{j\in S}w_{ij}=\sum_{j\in S}(1-W_{j}),
}
and similarly,
\eq{\label{eq: sum expanded 2}
	w_{\text{O}}(S)-w_{\text{I}}(S)=\sum_{i\in V}\sum_{j\not\in S}w_{ij}-\sum_{i\not\in S}\sum_{j\in V}w_{ij}=\sum_{j\not\in S}(W_{j}-1).
}
Thus Lemma \ref{lem: column bounds} implies
\eq{
	\absa{w_{\text{O}}(S)-w_{\text{I}}(S)}\leq \min \ha{k,n-k}\cdot \frac{C\pa{\log n}^{C+C\xi}}{\sqrt{n}},
}
for all $S$ with high probability on $\Omega$.

Lemma \ref{lem: weight bounds} implies
\eq{
	|w_{\text{I}}(S)| \geq c \min\ha{|S|,n-|S|}
}
for all $S$ with high probability on $\Omega$. Putting this together we see
\eq{\label{eq: in out bound}
	\absa{\frac{w_{\text{O}}(S)}{w_{\text{I}}(S)}-1}\leq \frac{C\pa{\log n}^{C+C\xi}}{\sqrt{n}}
}
for all $S$ with high probability on $\Omega$. However, by Lemma \ref{lem: Omega is hp}, the event $\Omega$ holds with high probability itself and thus unconditionally \eqref{eq: in out bound} holds for all $S$ with high probability.

Finally, applying Theorem \ref{thm: robust isothermal}, we get
\eq{
	\sup_{r>0}\absa{\rho_{G_{n}}(r)-\rhoM(r)}\leq \frac{C\pa{\log n}^{C+C\xi}}{\sqrt{n}}
}
with high probability.
\end{proof}

\begin{remark}[On Theorem \ref{thm: random graph}]
	The parameter $p$, the probability that an edge of some weight exists between two directed vertices, can be interpreted as a measure of the sparseness of the population structure. We can ask how few interactions on average can individuals in a population have with others and still yield populations with Moran-type behavior? While $p$ can be arbitrarily small, we have kept it constant and, in particular, not dependent on $n$. However, could $p$ depend on $n$ such that $p\to0$ as $n\to\infty$ and still produce graphs which show Moran-type behavior? An obvious lower bound on the rate of $p$'s convergence to $0$ is provided by the Erd\H{o}s-R\'{e}nyi model, which tells us that a graph is almost surely disconnected in the limit for $\sqrt{p}<(1-\e)(1/n)\log n$ for any $\e>0$. This bound follows by noting that $(1-p)^{2}$ is the probability that there is no edge, in either direction, between two vertices and then applying the usual Erd\H{o}s-R\'{e}nyi threshold \cite{Erdos1960, Chung2006}. There is much room between this lower bound and $p$ constant---even whether such a sharp threshold for $p$ exists is currently unclear. The issue is difficult to approach with naive simulations as the Moran process is not guaranteed to terminate on disconnected graphs. 
\end{remark}


\section{Graphs without outgoing weights summing to 1}

The Moran process on graphs may be generalized to no longer require the sum of the outing going weights of each vertex to be 1. The process is still defined for a directed, weighted graph, $G\equiv G_{n}=(V_{n},W_{n})$, where $V\equiv V_{n}\deq\db{n}$ but $W\equiv W_{n}\deq [w_{ij}]$ need not be a stochastic matrix  (it must still have nonnegative entries). Instead of sampling \emph{vertices} proportional to fitness and then choosing an outgoing edge with probability equal to its weight, we rather sample \emph{edges} proportional to their weights and the fitness of the individual at the beginning of the edge. Once we select an edge the type of the individual at the end of the edge becomes the same as the individual at the beginning of the edge. It is easy to see that we get the original process as a special case of the new model. Again, the process $X_{t}$ is Markovian with state space $2^{V}$, where each $X_{t}$ is the subset of $V$ consisting of all the vertices occupied by mutants at time $t$. At time 0 a mutant is placed at one of the vertices uniformly at random. We then update as described above by choosing edges proportional to their weights and the individual's type at the edges origin. For example, the probability of choosing a particular edge $(i,j)$ is
\eq{
	\frac{( r\I_{i\in X_{t}} + \I_{i\not\in X_{t}} )w_{ij}}{ r\sum_{i\in X_{t}}\sum_{j\in V}w_{ij}+\sum_{i\not\in X_{t}}\sum_{j\in V}w_{ij} }.
}

In this setting the robust isothermal theorem still holds.

\begin{theorem}[Robust isothermal theorem]\label{thm: robust isothermal 2}
	Fix $0\leq \e<1$. Let $G_{n}=(V_{n},W_{n})$ be a connected graph. If for all nonempty $ S\subsetneq V_{n}$ we have
	\eq{\label{eq: robust assumption 2}
		\absa{\frac{w_{\text{O}}(S)}{w_{\text{I}}(S)}-1}\leq \e,
	}
	then
	\eq{\label{eq: robust isothermal 2}
		\sup_{r>0}\absa{\rhoM(r) - \rho_{G_{n}}(r)} \leq  \e.
	}
\end{theorem}

\begin{proof}
	Just as before, we make the projection of the state space of all subsets of $V$, which records exactly which vertices are mutants, to the simpler state space $\{0,1,\ldots,n\}$, which records only the number of mutants. In the new model, we still have
	\eq{\
		\frac{ p_{+}(S) }{ p_{-}(S) }=r\frac{ w_{\text{O}}(S) }{ w_{\text{I}}(S) }.
	}
So the argument is exactly the same as the previous case.
\end{proof}

\section{More random graphs}\label{sec: random graphs 2}

We can introduce a random graph model where the sum of the outgoing weights are not equal, that is, where individuals can contribute differentially to the next time point in a way not dependent on the genotype they carry. We change the model by not normalizing the outgoing edge weights. Following the Erd\H{o}s-R\'{e}nyi model, we produce a weighted, directed graph as follows: Consider an $n\times n$ matrix $X=[x_{ij}]$ with independent, identically distributed, nonnegative random variables for its entries. Since we do not need to normalize we take, $W=X$.

\begin{definition}[Generalized Erd\H{o}s-R\'{e}nyi random graphs]\label{def: random graph 2}
	Let $\mu$ be a nonnegative distribution (not depending on $n$) with subexponential decay such that if $X\sim\mu$
	\eq{\label{eq: distribution assumptions 2}
		\P[X>0]= p>0 ~ \text{and} ~ \P\qa{ X\geq x }\leq Ce^{-x^{c}}
	}
	for some positive constants $p$, $c$, and $C$ and all $x>0$. We denote the mean and standard deviation of $X$ by $\mu_{1}$ and $\sigma$ respectively. We generate a family of random graphs $G_{n}=(V_{n},X_{n}\equiv W_{n})$ from $\mu$, where $x_{ij}$ are independent and distributed according to $\mu$.
\end{definition}

Now we have a similar theorem to before but the proof is easier.

\begin{theorem}\label{thm: random graph 2}
	Let $\pa{G_{n}}_{n\geq 1}$ be a family of random graphs as in Definition \ref{def: random graph 2}. Then there are constants $C>0$ and $c>0$, not dependent on $n$, such that the fixation probability of a randomly placed mutant of fitness $r>0$ satisfies
	\eq{\label{eq: random graphs 1 2}
		\absa{\rho_{G_{n}}(r)-\rhoM(r)}\leq  \frac{C\pa{\log n}^{C+C\xi}}{\sqrt{n}}
	}
	uniformly in $r$ with probability greater than
	\eq{
		1- \exp\pa{-\nu \pa{\log n}^{1+\xi}},
	}
	for positive constants $\xi$ and $\nu$.
\end{theorem}

\subsection{Proof of Theorem \ref{thm: random graph 2}} 
We now use the subexponential decay assumption to understand the typical behavior of the random variables $x_{ij}$. 

\begin{definition}[Good events $\Omega$]\label{def: Omega 2}
	Let $\Omega$ be an $n$-dependent event such that the following hold:
	\al{
		\Omega\deq \bigcap_{i,j=1}^{n} \ha{ x_{ij}\leq C\pa{\log n}^{C}} \cap\ha{ G_{n} \text{ is connected} }.
	}
\end{definition}

\begin{lemma}\label{lem: Omega is hp 2}
	The event $\Omega$ holds with high probability.
\end{lemma}

\begin{proof}
Identical to the previous proof.
\end{proof}

Define the sum of the $j$th column as
\eq{
	W_{j}\deq\sum_{i=1}^{n}w_{ij}
}
and the sum of the $i$th row as
\eq{
	\tilde{W}_{i}\deq\sum_{j=1}^{n}w_{ij}.
}

\begin{lemma}\label{lem: column bounds 2}
	On $\Omega$, there are positive constants $c\equiv c_{\mu}$ and $C\equiv C_{\mu}$, not dependent on $n$, such that the following inequalities hold
	\eq{
		\absa{ W_{j} -\mu n}\leq C\pa{\log n}^{C+C\xi}\sqrt{n}
	}
	and
	\eq{
		\absa{ \tilde{W}_{i} -\mu n}\leq C\pa{\log n}^{C+C\xi}\sqrt{n}
	}
	for all $i,j\in V$, with probability at least
	\eq{
		1-e^{-\nu \pa{\log n}^{1+\xi}}.
	}
\end{lemma}
\begin{proof}
Consider $a_{i}=w_{ij}-\mu$ or $a_{i}=w_{ji}-\mu$ and apply Lemma \ref{lem: lde}. The claim follows immediately as there are $2n$ high probability events.
\end{proof}

\begin{lemma}\label{lem: weight bounds 2}
	On $\Omega$, for all $\emptyset\neq S\subsetneq V_{n}$ and some small constant $c\equiv c_{\mu}>0$, not dependent on $n$, we have the following bound
	\eq{
		|w_{\text{I}}(S)|=|w_{\text{O}}(S^{c})| \geq c_{\mu} |S|\pa{n-|S|}
	}
	with probability greater than
	\eq{
		1-e^{-\nu\pa{\log n}^{1+\xi}}.
	}
\end{lemma}
\begin{proof}
By assumption on the distribution $\mu$, we have $\P\qa{ x_{ij}>0 }=p>0$ and thus there is a constant $c>0$ such that $\P\qa{ x_{ij}\geq c }=p'>0$. Define $\beta_{ij}\deq\I\pa{ x_{ij}\geq c }$ which are independent Bernoulli random variables such that
\eq{\label{eq: bernoulli lower bound 2}
	\P[\beta_{ij}=1]=p'>0,
}
since the $x_{ij}$ are independent. Let $|S|=k$. By definition
\eq{
	w_{\text{I}}(S)=\sum_{i\not\in S}\sum_{j\in S}x_{ij} =w_{\text{O}}(S^{c}).
}
Using \eqref{eq: bernoulli lower bound},
\eq{
	\sum_{i\in S}\sum_{i\not\in S}x_{ij} \geq c\sum_{i\in S}\sum_{i\not\in S}\beta_{ij}.
}
Note that now this is a sum of $k(n-k)$ independent random variables. So for fixed $\emptyset\neq S\subsetneq V_{n}$, by the Chernoff bound, Lemma \ref{lem: mcb}, the event
\eq{
	A_{S}\deq \ha{ \sum_{i\in S}\sum_{i\not\in S}\beta_{ij}\leq (1-1/2)p' k(n-k) }
}
has probability less than
\eq{
	\exp\pa{-\frac{1}{8}p'k(n-k)}.
}
Thus, by the union bound
\al{\begin{split}
	\P\qa{ \bigcap_{\emptyset\neq S\subsetneq V_{n}} A_{S}^{c}}&=1-\P\qa{ \bigcup_{\emptyset\neq S\subsetneq V_{n}} A_{S}}\\
	&\geq 1- \sum_{\emptyset\neq S\subsetneq V_{n}}\P\qa{A_{S}}\\
	&= 1-\sum_{k=1}^{n-1}\binom{n}{k}\P\qa{A_{S}} \\
	&=1-\sum_{k=1}^{\floor{\log n}}\binom{n}{k}\P\qa{A_{S}}-\sum_{k=\floor{\log n}+1}^{n-\floor{\log n}-1}\binom{n}{k}\P\qa{A_{S}}-\sum_{n-\floor{\log n}}^{n-1}\binom{n}{k}\P\qa{A_{S}}\\
	&\geq1-2n^{\log n}\exp\pa{ -p'n/8 }-2^{n}\exp\pa{-p'n\pa{\log n}/8}\\
	&\geq 1-2\exp\pa{ -\pa{p'n/8-(\log n)^{2} }}-\exp\pa{ -\pa{p'\log n/8-\log 2}n }\\
	&\geq 1- \exp\pa{ -cp'n },
\end{split}}
for some $c>0$. Finally, note 
\eq{
	\exp(-cp'n) \leq e^{-\nu\pa{\log n}^{1+\xi}}
}
for an appropriate choice of $\xi$ and $\nu$.
\end{proof}

We now complete the proof of Theorem \ref{thm: random graph 2}.

\begin{proof}[Proof of Theorem \ref{thm: random graph 2}]
Again let $|S|=k$. We check that the assumptions of Theorem \ref{thm: robust isothermal 2} hold with high probability. Observe
\eq{
	\absa{\frac{w_{\text{O}}(S)}{w_{\text{I}}(S)}-1}=\absa{ \frac{w_{\text{O}}(S)-w_{\text{I}}(S)}{w_{\text{I}}(S)} }=  \frac{\absa{w_{\text{O}}(S)-w_{\text{I}}(S)}}{\absa{w_{\text{I}}(S)}}.
}
Expanding the numerator, we get
\eq{\label{eq: sum expanded 1 2}
	w_{\text{O}}(S)-w_{\text{I}}(S)=\sum_{i\in S}\sum_{j\not\in S} w_{ij} -\sum_{i\not\in S}\sum_{j\in S}w_{ij}=\sum_{i\in S}\sum_{j\in V}w_{ij}-\sum_{i\in V}\sum_{j\in S}w_{ij}=\sum_{j\in S}(\tilde{W}_{j}-W_{j}),
}
and similarly,
\eq{\label{eq: sum expanded 2 2}
	w_{\text{O}}(S)-w_{\text{I}}(S)=\sum_{i\in V}\sum_{j\not\in S}w_{ij}-\sum_{i\not\in S}\sum_{j\in V}w_{ij}=\sum_{j\not\in S}(W_{j}-\tilde{W}_{j}).
}
Thus Lemma \ref{lem: column bounds 2} implies
\eq{
	\absa{w_{\text{O}}(S)-w_{\text{I}}(S)}\leq \min \ha{k,n-k}\cdot C\pa{\log n}^{C+C\xi}\sqrt{n},
}
for all $S$ with high probability on $\Omega$.

Lemma \ref{lem: weight bounds 2} implies
\eq{
	|w_{\text{I}}(S)| \geq c k\pa{n-k}
}
for all $S$ with high probability on $\Omega$. Putting this together we see
\eq{\label{eq: in out bound 2}
	\absa{\frac{w_{\text{O}}(S)}{w_{\text{I}}(S)}-1}\leq \frac{C\pa{\log n}^{C+C\xi}}{\sqrt{n}}
}
for all $S$ with high probability on $\Omega$. However, by Lemma \ref{lem: Omega is hp 2}, the event $\Omega$ holds with high probability itself and thus unconditionally \eqref{eq: in out bound} holds for all $S$ with high probability.

Finally, applying Theorem \ref{thm: robust isothermal 2}, we get
\eq{
	\sup_{r>0}\absa{\rho_{G_{n}}(r)-\rhoM(r)}\leq \frac{C\pa{\log n}^{C+C\xi}}{\sqrt{n}}
}
with high probability.
\end{proof}


\section{Large deviation estimates and concentration inequalities}

In this section we provide a brief review of large deviation estimates and concentration inequalities with a focus on those used above. A large deviation estimate ({\sc lde}) controls atypical behavior of sums of independent (or sometimes weakly dependent) random variables, whereas a concentration inequality controls the convergence of an average of independent (or sometimes weakly dependent) random variables to their mean. For a more in-depth review of {\sc lde}s, see for example \cite{Chung2006, Ledoux2001}. Many {\sc lde}s follow directly by applying Markov's inequality, so we state this now.

\begin{theorem}[Markov's inequality]\label{thm: markov}
	Let $X$ be a nonnegative random variable and $t> 0$. Then
	\eq{
		\P[X\geq t]\leq \frac{\E X}{t}.
	}
\end{theorem}
\begin{proof}
	Define the indicator random variable $\I_{X\geq t}$. Then $t\I_{X\geq t}\leq X$, thus $\E[ t\I_{X\geq t}]\leq \E X$. Therefore,
	\eqs{
		\P[X\geq t]=\E \I_{X\geq t}\leq \frac{\E X}{t}.
	}
\end{proof}

This very simple result has lots of scope. The general idea is to define a nonnegative, increasing function $f$ of some random variable $X$ and note that Markov's inequality implies
\eq{
	\P[ X\geq t ]\leq \P[ f(X)\geq f(t) ]\leq \frac{\E f(X)}{f(t)}.
}
Normally, $f$ is chosen as $x^{k}$ or $e^{\lambda X}$ where $k$ or $\lambda>0$ is optimized to strengthen the inequality. If the random variable $X$ is a sum of centered, independent random variables, $\sum_{i=1}^{n}\pa{X_{i}-\E X_{i}}$, the function takes the form
\eq{
	\prod_{i=1}^{n}\exp\pa{ \lambda\pa{ X_{i}-\E X_{i} } }.
}
In this way we get several inequalities.

\begin{lemma}[Hoeffding's inequality]\label{lem: hi}
	Suppose that $X_{1},\ldots, X_{n}$ are i.i.d. Bernoulli random variables with parameter $p\in[0,1]$. Define $X\deq\sum_{i=1}^{n}X_{i}$. Then 
	\eq{
		\P\qa{ \absa{ X-\E X }\geq \delta\sqrt{n} }\leq 2\exp\pa{ -2\delta^{2} }
	}
	for all $\delta>0$.
\end{lemma}

Lemma \ref{lem: hi} states that $X$ fluctuates about its expectation on the order of $\sqrt{n}$, and the probability of a fluctuant greater than $\delta\sqrt{n}$ decays exponential with $\delta>0$. The next lemma bounds fluctuation of larger orders, and thus they occur even more infrequently. We shall only need a lower bound in this case:

\begin{lemma}[Multiplicative Chernoff bound]\label{lem: mcb}
	Suppose that $X_{1},\ldots, X_{n}$ are i.i.d. Bernoulli random variables with parameter $p\in[0,1]$. Define $X\deq\sum_{i=1}^{n}X_{i}$. Then 
	\eq{
		\P\qa{  X \leq (1-\e)\E X }\leq \exp\pa{ -\frac{\e^{2}p}{2}n }
	}
	and
	\eq{
		\P\qa{  X \geq (1+\e)\E X }\leq \exp\pa{ -\frac{\e^{2}p}{2}n }.
	}
\end{lemma}

We remark that far more general statements of Lemmas \ref{lem: hi} and \ref{lem: mcb} are possible, but we state only the versions we use in Section \ref{sec: random graphs}.

Finally, we state a {\sc lde} for weighted sums of independent random variables with the following conditions on their moments:
\eq{\label{eq: lde moment conditions}
	\E X=0, \quad \E\absa{X}^{2}=\sigma^{2}, \quad \text{and} \quad   \E \absa{X}^{k}  \leq (Ck)^{Ck},
}
for some positive constant $C>0$ (not dependent on $n$ or $k$) and for $k\geq 1$.

\begin{lemma}\label{lem: lde}
	Suppose the independent random variables $\pa{a_{i}^{(n)}}_{i=1}^{n}$ for $n\in\N$ satisfy \eqref{eq: lde moment conditions} and that $\pa{A_{i}^{(n)}}_{i=1}^{n}$ for $n\in\N$ are constants in $\R$. Then
	\eq{
		\P\qa{ \absa{ \sum_{i=1}^{n} a_{i}A_{i}  } \geq \sigma\pa{\log n}^{C+C\xi} \pa{\sum_{i=1}^{n} \absa{A_{i}}^{2} }^{1/2} } \leq e^{-\nu\pa{\log n}^{1+\xi}}.
	}
	In words, we can bound the sum $\sum_{i=1}^{n} a_{i}A_{i}$ on the same order as the norm of the coefficients with high probability.
\end{lemma}

To prove this lemma we use a high-moment Markov inequality, so first we need a result bounding the higher moments of this sum.

\begin{lemma}\label{lem: high moment bound}
	Suppose the independent random variables $\pa{a_{i}^{(n)}}_{i=1}^{n}$ for $n\in\N$ satisfy \eqref{eq: lde moment conditions} and that $\pa{A_{i}^{(n)}}_{i=1}^{n}$ for $n\in\N$ are constants in $\R$. Then
	\eq{
		\E\absa{ \sum_{i=1}^{n} a_{i}A_{i}  }^{k}  \leq  \pa{Ck}^{Ck} \pa{\sum_{i=1}^{n} \absa{A_{i}}^{2} }^{k/2}
	}
\end{lemma}

\begin{proof}
Without loss of generality let $\sigma=1$. Let $A^{2}\deq \sum_{i}\absa{A_{i}}^{2}$, then by the classical Marcinkiewicz-Zygmund inequality \cite{Stroock2011a} in the first line, we get
\al{
	\E\absa{ \sum_{i} a_{i}A_{i}  }^{k}  &\leq (Ck)^{k/2} \E\absa{  \pa{\sum_{i}\absa{A_{i}}^{2}\absa{a_{i}}^{2} }^{1/2} }^{k} \\
	& = (Ck)^{k/2} A^{k} \E \qa{ \pa{\sum_{i}\frac{\absa{A_{i}}^{2} }{A^{2}}\absa{a_{i}}^{2} }^{k/2}  } \\
	& \leq (Ck)^{k/2} A^{k} \E\qa{ \sum_{i}\frac{\absa{A_{i}}^{2}}{A^{2}}\absa{a_{i}}^{k} } \\
	&= (Ck)^{k/2} A^{k} \sum_{i} \frac{\absa{A_{i}}^{2}}{A^{2}} \E\absa{ a_{i} }^{k} \\
	&\leq (Ck)^{Ck+k/2} A^{k}\\
	&\leq (Ck)^{Ck} A^{k},
}
where we have used Jensen's inequality in the third line and assumption \eqref{eq: lde moment conditions} in line 5.
\end{proof}

\begin{proof}[Proof of Lemma \ref{lem: lde}]
	Without loss of generality let $\sigma=1$. The proof is a simple application of Markov's inequality, Theorem \ref{thm: markov}. Let $k=\nu \pa{ \log n }^{1+\xi}$, then by Lemma \ref{lem: high moment bound}, we get
	\al{
		\P\qa{ \absa{ \sum_{i} a_{i}A_{i}  } \geq \pa{\log n}^{C+C\xi} \pa{\sum_{i} \absa{A_{i}}^{2} }^{1/2} } & = \P\qa{ \absa{ \sum_{i} a_{i}A_{i}  }^{k} \geq \pa{\log n}^{Ck+Ck\xi} \pa{\sum_{i} \absa{A_{i}}^{2} }^{k/2} } \\
		&\leq  \frac{\E\absa{ \sum_{i} a_{i}A_{i}  }^{k}}{\pa{\log n}^{Ck+Ck\xi} \pa{\sum_{i} \absa{A_{i}}^{2} }^{k/2}} \\
		&\leq \pa{\frac{ Ck }{\pa{\log n}^{1+\xi}  } }^{Ck} \\
		&= \pa{ C\nu }^{C\nu \pa{ \log n }^{1+\xi}} \\
		&\leq e^{-\nu (\log n)^{1+\xi}},
	}
	for $\nu\leq e^{-1}$ small enough.
\end{proof}

\biblio


\end{document}